\documentclass[twocolumn]{aastex63}

\hypersetup{linkcolor=blue,citecolor=blue,filecolor=cyan,urlcolor=blue}

\usepackage{CJK}

\usepackage{longtable}
\usepackage{lipsum} 
\usepackage{multirow}

\usepackage{natbib}
\usepackage{graphicx}
\usepackage{bm}
\usepackage{amsmath}
\usepackage{amsfonts}
\usepackage{amssymb}
\usepackage{xcolor}
\usepackage{mathrsfs}
\usepackage[caption=false]{subfig}
\usepackage{CJK}
\usepackage{ulem}

\shorttitle{NGC\,6334S}
\shortauthors{Li et al.}
\usepackage{CJK}

\graphicspath{{./}{figures/}}
\usepackage{xspace}
\newcommand{\um}{{$\mu$m}\xspace}
\def\arcsec{$^{\prime\prime}$\xspace}
\def\Mo{M$_{\odot}$\xspace}
\def\kms{km s$^{-1}$\xspace}

\def\1{\uppercase\expandafter{\romannumeral1}}
\def\2{\uppercase\expandafter{\romannumeral2}}
\def\3{\uppercase\expandafter{\romannumeral3}}
\def\4{\uppercase\expandafter{\romannumeral5}}

\newcommand{\x}{\textcolor{red}{XX}}

\begin{document}

\title{ALMA observations of NGC\,6334S. \uppercase\expandafter{\romannumeral2}. 
Subsonic and Transonic Narrow Filaments in a High-mass Star Formation Cloud}

\correspondingauthor{Shanghuo Li}
\email{shanghuo.li@gmail.com}

\author[0000-0003-1275-5251]{Shanghuo Li }
\affiliation{Korea Astronomy and Space Science Institute, 776 Daedeokdae-ro, Yuseong-gu, Daejeon 34055, Republic of Korea}

\author[0000-0002-7125-7685]{Patricio Sanhueza}
\affiliation{National Astronomical Observatory of Japan, National Institutes of Natural Sciences, 2-21-1 Osawa, Mitaka, Tokyo 181-8588, Japan}
\affiliation{Department of Astronomical Science, SOKENDAI (The Graduate University for Advanced Studies), 2-21-1 Osawa, Mitaka, Tokyo 181-8588, Japan}

\author[0000-0002-3179-6334]{Chang Won Lee}
\affil{Korea Astronomy and Space Science Institute, 776 Daedeokdae-ro, Yuseong-gu, Daejeon 34055, Republic of Korea}
\affil{University of Science and Technology, 217 Gajeong-ro, Yuseong-gu, Daejeon 34113, Republic of Korea}

\author[0000-0003-2384-6589]{Qizhou Zhang}
\affiliation{Center for Astrophysics $|$ Harvard \& Smithsonian, 60 Garden Street, Cambridge, MA 02138, USA}

\author[0000-0002-1700-090X]{Henrik Beuther}
\affiliation{Max Planck Institute for Astronomy, Konigstuhl 17, 69117 Heidelberg, Germany}

\author[0000-0002-9569-9234]{Aina Palau}
\affiliation{Instituto de Radioastronom\'ia y Astrof\'isica, Universidad Nacional Aut\'onoma de M\'exico, P.O. Box 3-72, 58090, Morelia, Michoac\'an, M\'exico}

\author{Hong-Li Liu}
\affiliation{Department of Astronomy, Yunnan University, Kunming, 650091, People's Republic of China}

\author{Howard Smith}
\affiliation{Center for Astrophysics $|$ Harvard \& Smithsonian, 60 Garden Street, Cambridge, MA 02138, USA}

\author[0000-0003-2300-2626]{Hauyu Baobab Liu}
\affiliation{Academia Sinica Institute of Astronomy and Astrophysics, 11F of AS/NTU Astronomy-Mathematics Building, No.1, Sec. 4, Roosevelt Road, Taipei 10617, Taiwan}

\author[0000-0003-4493-8714]{Izaskun, Jim\'enez-Serra}
\affiliation{Centro de Astrobiolog\'ia (CSIC-INTA), Carretera de Ajalvir, Km. 4, Torrej\'on de Ardoz, 28850 Madrid, Spain}

\author[0000-0003-2412-7092]{Kee-Tae Kim}
\affil{Korea Astronomy and Space Science Institute, 776 Daedeokdae-ro, Yuseong-gu, Daejeon 34055, Republic of Korea}
\affil{University of Science and Technology, 217 Gajeong-ro, Yuseong-gu, Daejeon 34113, Republic of Korea}

\author[0000-0002-4707-8409]{Siyi Feng}
\affiliation{National Astronomical Observatories, Chinese Academy of Sciences, Beijing 100101, People's Republic of China}
\affiliation{Academia Sinica Institute of Astronomy and Astrophysics,11F of AS/NTU Astronomy-Mathematics Building, No.1, Sec. 4, Roosevelt Road, Taipei 10617, Taiwan}
\affiliation{National Astronomical Observatory of Japan, 2-21-1 Osawa, Mitaka, Tokyo, 181-8588, Japan}

\author[0000-0002-5286-2564]{Tie Liu}
\affiliation{Shanghai Astronomical Observatory, Chinese Academy of Sciences, 80 Nandan Road, Shanghai 200030, People's Republic of China}

\author[0000-0001-6106-1171]{Junzhi Wang}
\affil{Shanghai Astronomical Observatory, Chinese Academy of Sciences, 80 Nandan Road, Shanghai 200030, People's Republic of China} 

\author[0000-0003-3010-7661]{Di Li}
\affiliation{National Astronomical Observatories, Chinese Academy of Sciences, Beijing 100101, People's Republic of China}
\affil{NAOC-UKZN Computational Astrophysics Centre, University of KwaZulu-Natal, Durban 4000, South Africa}

\author[0000-0002-5093-5088]{Keping Qiu}
\affiliation{School of Astronomy and Space Science, Nanjing University, 163 Xianlin Avenue, Nanjing 210023, People's Republic of China}

\author[0000-0003-2619-9305]{Xing Lu}
\affil{Shanghai Astronomical Observatory, Chinese Academy of Sciences, 80 Nandan Road, Shanghai 200030, People's Republic of China} 
\affiliation{National Astronomical Observatory of Japan, National Institutes of Natural Sciences, 2-21-1 Osawa, Mitaka, Tokyo 181-8588, Japan}

\author[0000-0002-3829-5591]{Josep Miquel. Girart}
\affiliation{Institut de Ci\`encies de l'Espai (IEEC-CSIC), Campus UAB, Carrer de Can Magrans s/n, 08193 Cerdanyola del Vall\`es, Catalonia, Spain} 

\author[0000-0002-7237-3856]{Ke Wang}
\affil{Kavli Institute for Astronomy and Astrophysics, Peking University, 5 Yiheyuan Road, Haidian District, Beijing 100871, People's Republic of China}

\author[0000-0002-9832-8295]{Fei Li} 
\affiliation{School of Astronomy and Space Science, Nanjing University, 163 Xianlin Avenue, Nanjing 210023, People's Republic of China}

\author[0000-0003-3520-6191]{Juan Li}
\affil{Shanghai Astronomical Observatory, Chinese Academy of Sciences, 80 Nandan Road, Shanghai 200030, People's Republic of China}

\author{Yue Cao} 
\affiliation{School of Astronomy and Space Science, Nanjing University, 163 Xianlin Avenue, Nanjing 210023, People's Republic of China}
\affiliation{Center for Astrophysics $|$ Harvard \& Smithsonian, 60 Garden Street, Cambridge, MA 02138, USA}

\author[0000-0001-9333-5608]{Shinyoung Kim}
\affil{Korea Astronomy and Space Science Institute, 776 Daedeokdae-ro, Yuseong-gu, Daejeon 34055, Republic of Korea}
\affil{University of Science and Technology, 217 Gajeong-ro, Yuseong-gu, Daejeon 34113, Republic of Korea}

\author{Shaye Strom}
\affiliation{Center for Astrophysics $|$ Harvard \& Smithsonian, 60 Garden Street, Cambridge, MA 02138, USA}

\begin{abstract}
We present a study of narrow filaments toward a massive infrared 
dark cloud, NGC\,6334S, using the Atacama Large 
Millimeter/submillimeter Array (ALMA). 
Thirteen gas filaments are identified using the  H$^{13}$CO$^{+}$ 
line, while a single continuum filament is revealed by the 
continuum emission. 
The filaments present a compact radial distribution with a 
median filament width of $\sim$0.04 pc  
narrower than the previously proposed `quasi-universal' 
0.1~pc filament width. The higher spatial resolution 
observations and higher-density gas tracer tend to 
identify even narrower and lower mass filaments.  The filament 
widths are roughly twice the size of embedded cores.  
The gas filaments are largely supported by thermal motions. 
The nonthermal motions are predominantly subsonic and 
transonic in both identified gas filaments and embedded cores, 
which may imply that stars are likely born in environments of 
low turbulence. 
A fraction of embedded objects show a narrower velocity 
dispersion  compared with their corresponding 
natal filaments, 
which may indicate that the turbulent dissipation   
is taking place in these embedded cores. 
The physical properties (mass, mass per unit length, gas 
kinematics, and width) of gas filaments are analogous to those 
of narrow filaments found in low- to high-mass star-forming regions.  
The more evolved sources are found to be farther away from the 
filaments, a situation that  may have resulted from the relative 
motions between the YSOs and their natal filaments.

\end{abstract}



\section{Introduction} 
\label{sec:intro}
Filamentary structures of interstellar medium (ISM) are 
prevalent in nearby Gould Belt molecular clouds and also more 
distant molecular clouds as seen in recent Galactic plane surveys 
from far-infrared to centimeter wavelengths and in both continuum 
and molecular line emission 
\citep{2009PASP..121..213C,2010A&A...518L.102A,
2010A&A...518L.100M,2011A&A...529L...6A,2014ApJ...797...53G,
2015MNRAS.450.4043W,2015ApJ...815...23Z,2016MNRAS.456.2041C,
2016A&A...591A...5L,2020A&A...642A.163S,2020A&A...641A..53W}. 
These filaments show wide ranges of physical properties (e.g., 
length, width,  mass, length-to-width aspect ratios, and masses 
per unit length) that can vary over an order of magnitude across the 
revealed filaments.

Similar filamentary structures are also commonly 
seen in both numerical hydrodynamic (HD) and magnetohydrodynamic 
(MHD) simulations of the ISM \citep[e.g.,][]{2007ApJ...661..972P,
2008ApJ...674..316H,2011ApJ...729..120G,
2013A&A...556A.153H,2014ApJ...791..124G}.  
Several mechanisms have been proposed for the formation of 
filaments in molecular clouds, such as gravitational instability 
(gravitational fragmentation and collapse) of sheet-like and 
elongated clouds \citep{1987PThPh..78.1051M,1998ApJ...506..306N, 
2007ApJ...654..988H,2013A&A...556A.153H,2014ApJ...791..124G,
2014ApJ...789...37V}, 
cloud collision \citep{2001ApJ...553..227P}, 
and shocked flows \citep{2011ApJ...729..120G,
2020MNRAS.494.3675C}.

In dense and self-gravitating clouds,  filaments often 
exhibit cylindrical morphologies. 
\citep[e.g.,Taurus B213;][]{2012ApJ...756...12L}. 
Large scale filaments often harbor parsec-scale dense massive 
clumps that become the fertile ground of massive star and 
cluster formation \citep{2009ApJ...696..268Z,2014MNRAS.439.1996J,
2014MNRAS.439.3275W,2016ApJ...819..139B}, although not all 
filaments show signs of star-formation activity 
\citep[e.g., only pre-stellar cores are detected in 
Polaris flare;][]{2010A&A...518L.104M}. 
The embedded dense cores that are precursors of stars 
can be formed in the highest density regions of the filament 
via contraction by self-gravity and local kinematic processes 
\citep{1992ApJ...388..392I,2007ApJ...654..988H,
2008ApJ...674..316H,2009ApJ...704.1735H,
2008ApJ...687..354N,2009ApJ...700.1609M,
2011ApJ...729..120G}. 
The prestellar cores and protostellar cores 
are primarily found to reside in dense filamentary 
structures with supercritical mass per unit length in 
both low- and high-mass star-forming  molecular 
clouds \citep{2014prpl.conf...27A,2019ApJ...877..114C,
2019A&A...629A..81T}, 
and most of them  are believed to have formed by 
cloud collapse and/or fragmentation along filaments  
\citep{2010A&A...518L.103M,2014prpl.conf...27A,
2014MNRAS.440.2860H,
2014A&A...561A..83P,2015A&A...584A..67B,
2015A&A...584A..91K,2017MNRAS.468.2489C}. 
The gas flows along filaments can continuously supply 
the material for cores to grow in mass 
\citep{2012ApJ...756...10L,2013ApJ...766..115K,
2017ApJ...840...22L,2018ApJ...852...12Y,
2018ApJ...855....9L,2019MNRAS.487.1259L,
2019A&A...629A..81T,2021ApJ...915L..10S}.

Recently, ALMA high angular resolution observations reveal  
that narrow (i.e., filament widths of a few 0.01 pc) filamentary 
structures \citep[or ``fibers" in][]{2018A&A...610A..77H}, 
are found in some  high-mass star-forming clouds 
\citep[e.g., Orion, G035.39-00.33, and G14.225-0.506;][and 
references therein]{2014MNRAS.440.2860H,2018A&A...610A..77H,
2018ApJ...861...77M,2019ApJ...875...24C}.
These filaments are much narrower than the `quasi-universal' 
0.1~pc filament width 
proposed by previous studies using \textit{Herschel} observations 
\citep[e.g.,][and references therein]{2014prpl.conf...27A,
2019A&A...621A..42A}, and appear to  be 
intimately linked to dense cores \citep{2018A&A...610A..77H}. 
However,  whether such narrow filamentary structures are 
ubiquitous in high-mass star formation clouds, and what their 
properties are remain controversial topics to be more fully 
explored.

\begin{figure*}[!ht]
\centering
\includegraphics[scale=0.51]{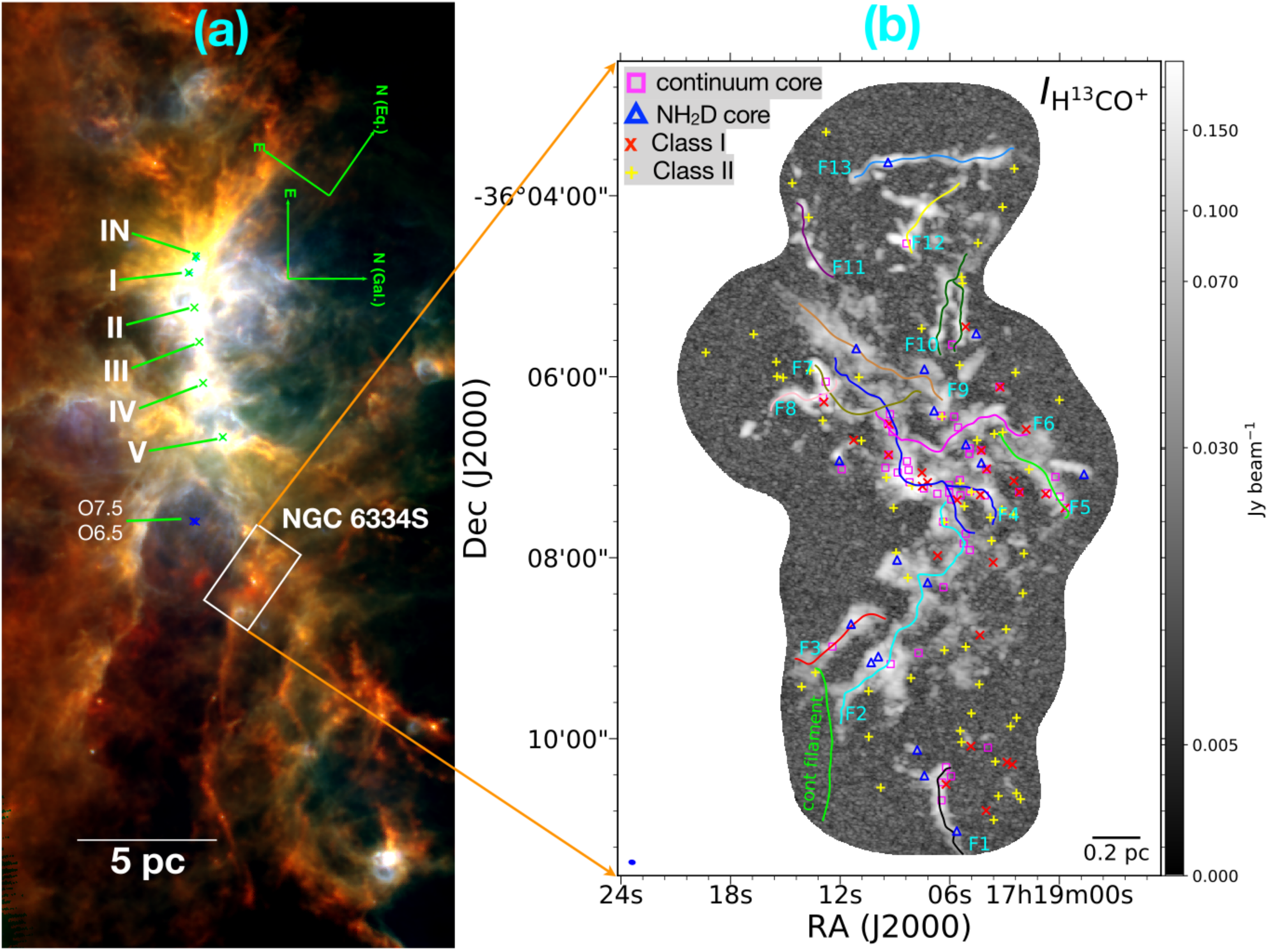}
\caption{
Panel a: three-color \textit{Herschel} composite image of NGC 6334 
molecular cloud complex with blue, green, and red for $\lambda$ = 
70, 160, and 350 \um,  respectively. The scale bar (5 pc at the 
distance of 1.3 kpc), the Equatorial and the Galactic cardinal 
directions are shown on the upper right hand of the image. 
The white box presents the NGC\,6334S region. Six bright 
infrared (IR) clumps (I, IN, II, III, IV, and V) are marked in the 
image \citep{1979ApJ...232L.183M}. Two O type stars (O7.5 
and O6.5) are marked with blue cross ``x"  symbols 
\citep{2008hsf2.book..456P}.
Panel b: the filament spines (color solid curves) overlaid on 
the peak intensity ($I_{\rm H^{13}CO^{+}}$; the maximum intensity 
of the spectrum) image of H$^{13}$CO$^{+}$ line 
emission. 
Magenta open squares correspond to the 49 continuum cores 
identified by the ALMA~3mm  continuum image (Paper~I). 
Blue open triangles show the 17 NH$_{2}$D cores revealed by the 
NH$_{2}$D line emission (Paper~II). 
The red cross ``x" and yellow plus ``+" symbols correspond to 
the 25 Class I and 58 Class II YSOs \citep{2013ApJ...778...96W}, 
respectively. 
The beam size (blue filled ellipse) of the H$^{13}$CO$^{+}$ image 
is shown in the bottom left of the panel. 
}
 \label{fig:rgb}
\end{figure*}

To understand the nature of filaments and embedded dense cores 
in massive star formation regions,  we have carried out high angular 
resolution observations toward a filamentary infrared dark cloud,  
NGC\,6334S, using the Atacama Large Millimeter/submillimeter 
Array (ALMA).  
NGC\,6334S is located at the southwestern end of the NGC 6334 
molecular cloud complex  (Figure~\ref{fig:rgb}), which is a 
nearby (1.3 kpc) young and massive ``mini-starburst" region 
\citep{2014ApJ...784..114C,2013ApJ...778...96W}.  
In contrast to the well-known infrared bright OB cluster-forming 
clumps NGC\,6334I/I(N)/II/III/IV/V \citep{2008hsf2.book..456P,
2013A&A...554A..42R}, NGC\,6334S in some areas is dark in the 
infrared at wavelengths up to 70 \um   
\citep[see Figure~1 of ][hereafter Paper~I]{2020ApJ...896..110L}, 
signalling its youth. NGC\,6334S has a mass of $\sim$1390 \Mo 
(Paper~I), which is comparable to the clumps with embedded 
massive protostars and protoclusters elsewhere in the complex, 
and therefore has the potential to form massive stars 
together with lower-mass star clusters. 
Thus NGC\,6334S 
provides an ideal laboratory to investigate the early 
evolutionary stages of cluster formation in  filamentary clouds.  
We will use dense gas tracers and continuum emission 
not only to identify the filamentary structures in the 
position-position-velocity (PPV) space but also to study  the 
physical properties (e.g., gas kinematics, mass, structure profile) 
of both filaments and dense cores,  in order to understand 
the initial cloud environment of filament-based cluster formation.

We recently identified 49 continuum dense cores (hereafter 
continuum cores, named respectively \#1, \#2, \#3 ...) 
using our 3~mm continuum image (Paper~I) and found 17 
starless cores (hereafter NH$_{2}$D cores, namely M1, 
M2, M3 ...) using the NH$_{2}$D line emission  
\citep[][hereafter Paper~II]{2021ApJ...912L...7L}.  
These NH$_{2}$D cores are neither associated with continuum 
cores nor with Class \1/\2 young stellar objects 
\citep[YSOs;][]{2013ApJ...778...96W}. 
For simplicity, we refer to continuum cores and NH$_{2}$D 
cores as dense cores. 
The derived masses of dense cores range from 0.13 to 14.1~\Mo, 
with the mean and median values of 1.8 and 0.8~\Mo, respectively.  
The sizes of dense cores are between 0.01 and 0.04~pc, with the 
mean and median values of 0.018 and 0.017~pc, respectively.  
Paper~I also shows that the 
nonthermal motions are predominantly subsonic and transonic 
throughout NGC\,6334S and that the external pressure is important in 
confining the embedded objects. Paper~II reported the presence of 
a cluster of low-mass starless and pre-stellar cores that  show 
small velocity dispersions, a high fractional abundance of NH$_{2}$D, 
high NH$_{3}$ deuterium fractionation, and are dark at infrared 
wavelengths to 70~$\mu$m. In at least some of the NH$_{2}$D 
cores, turbulence seems dissipated and the gas kinematics is 
dominated by thermal motions.  

In this work, we focus on filaments and investigate their properties 
as well as the relationship between filaments and dense 
cores. The observations are described in Section~\ref{sec:obs}. 
Then, we describe the filament identification and the properties of 
identified filaments in Section~\ref{sec:res}. We discuss in detail 
the properties of filaments and dense objects in Section~\ref{sec:dis}.  
Finally, we summarize our main findings in Section~\ref{sec:con}.

\section{Observation} 
\label{sec:obs}
We have carried out a 55-pointings mosaic observation 
with ALMA 12m array towards the massive infrared dark 
cloud (IRDC) NGC\,6334S  between March 
13 and 21 of 2017 (ID: 2016.1.00951.S).  
Two 234.4 MHz wide spectral windows 
were employed to cover the H$^{13}$CO$^{+}$ (1-0, 86.7 GHz) 
and NH$_{2}$D ($1_{1,1}-1_{0,1}$, 85.9 GHz) lines with a 
0.061 MHz spectral resolution ($\sim$0.21 km\,s$^{-1}$ 
at 86 GHz).  In addition, three 1.875 GHz wide spectral 
windows centered at 88.5 GHz, 98.5 GHz, and 100.3 GHz with 
a spectral resolution of 0.977 MHz  ($\sim$3.0 -- 3.3 km\,s$^{-1}$) 
were used to take broad band continuum data. More details on 
the observations can be found in Paper~I. 

Data calibration was performed using the CASA 4.7.0 software 
package \citep{2007ASPC..376..127M}.   
Both continuum and line images were iteratively cleaned 
with manual masking via the \textit{clean} task down to 
$\sim$3$\sigma$ using the multiscale deconvolver and 
a robust weighting of 0.5.  
The resultant continuum and line images have a 
 synthesized beam of 
$\theta_{\rm maj} \, \times \, \theta_{\rm min}$  = 
3\arcsec.6 $\times$ 2\arcsec.4  (or 0.023 $\times$ 0.015 pc, 
with a position angle P.A  = 81$^{\circ}$)  and  
$\theta_{\rm maj} \, \times \, \theta_{\rm min}$  = 
4\arcsec.1 $\times$ 2\arcsec.8 (or 0.026 $\times$ 0.018 pc, 
P.A  = 83$^{\circ}$), respectively.  
The achieved 1$\sigma$ root mean square (rms) 
noise levels are  0.3 mJy\,beam$^{-1}$ for the continuum 
image and $\sim$6 mJy\,beam$^{-1}$ per 0.21~km\,s$^{-1}$ 
for the spectral line images.  The maximum recoverable 
scale (MRS) of single pointing reaches up to 
$\sim$30\arcsec\ in the ALMA data. 
All images shown in this paper are prior to primary 
beam correction. The measured fluxes for mass 
estimation have the primary beam correction applied.

\begin{figure*}[ht!]
\epsscale{1.19}
\plotone{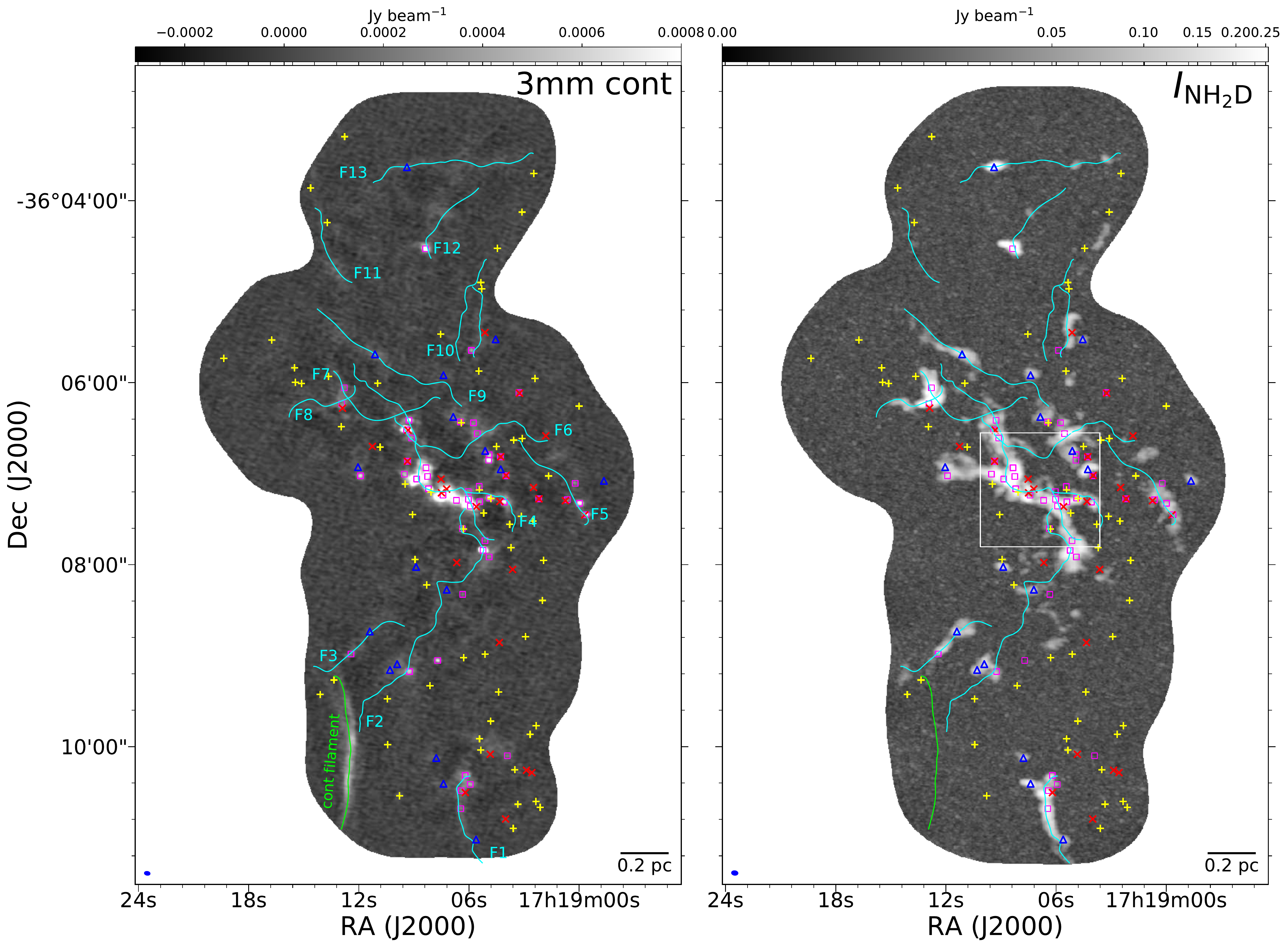}
\caption{Magenta open squares, blue open triangles, red cross ``x", and 
yellow plus ``+" symbols show the continuum cores, NH$_{2}$D cores, 
Class~\1, and Class~\2, respectively. 
Left: the filament spines (cyan and green solid curves)  
overlaid on the 3~mm continuum image. 
Right: the filament spines (cyan and green solid curves) overlaid 
on the peak intensity map ($I_{\rm NH_{2}D}$) of the NH$_{2}$D line 
emission.  White box shows the region where outflows are 
identified, with zoomed-in views presented in Figures~\ref{fig:outflow}. 
The beam size is shown in the bottom left of each panel.
 \label{fig:cont}}
 \end{figure*}

\section{Results and Analysis} 
\label{sec:res}
NGC\,6334S is mostly dark at infrared wavelengths to 
70~\um indicating its early evolutionary stage. \citep[e.g.,][]{2013ApJ...773..123S,2019ApJ...886..102S,
2013ApJ...779...96T,2018ApJ...861...14C,
2017ApJ...841...97S,2019ApJ...886..130L,
2021arXiv210901231M}. 
Figure~\ref{fig:rgb} shows an overview of 
the NGC\,6334 molecular cloud complex in the far-infrared 
and the location of NGC\,6334S.

\subsection{Molecular Lines Emission} 
\label{sec:oview}
The rotational transitions of several molecular species (i.e., 
HCO$^{+}$ (1--0), HCN (1--0), CS (2--1), HNCO ($4_{0,4}-3_{0,3}$), 
H$^{15}$NC (1--0), CH$_{3}$OH ($5_{1,4}-4_{1,3}$), SO ($2_{2}-1_{1}$), 
HC$_{3}$N (11--10)) were detected with a coarse spectral resolution 
of 0.977 MHz (or $\sim$3.0 -- 3.3 \kms). However  with these low  
spectral resolution data we are not able to determine  
the kinematic properties of the molecular gas.  Therefore, only 
the high spectral resolution (0.061 MHz $\sim$ 0.21 \kms) data 
of the H$^{13}$CO$^{+}$ (1-0) and NH$_{2}$D ($1_{1,1}-1_{0,1}$) 
lines will be used as diagnostics of the kinematic  
properties of the filaments in this work.

Figures~\ref{fig:rgb} and \ref{fig:cont} show the 
H$^{13}$CO$^{+}$ line, continuum,  and NH$_{2}$D line 
emission. The H$^{13}$CO$^{+}$ (1-0; critical density 
$n_{\rm cr} \, \sim$ 10$^{5}$ cm$^{-3}$) line 
traces spatially much more extended gaseous structures than 
the NH$_{2}$D ($1_{1,1}-1_{0,1}$; $n_{\rm cr} \, 
\sim$ 10$^{6}$ cm$^{-3}$) line since their critical 
densities are different by nearly an order-of-magnitude. 
The NH$_{2}$D 
emission appears preferentially toward the location of dense 
cores.   In addition, the H$^{13}$CO$^{+}$ emission is in a 
better agreement  with the \textit{Spitzer} dark and 
\textit{Hershcel} bright filamentary structures than that of 
NH$_{2}$D (see also Paper~I). 
These all suggest that the H$^{13}$CO$^{+}$ emission is a better 
 tracer of filamentary structures than NH$_{2}$D. 
In what follows, the H$^{13}$CO$^{+}$ will be therefore used to 
identify the  velocity-coherent filamentary structures. 
There is the continuum filamentary structure in the south-eastern 
part of the map (see Figures \ref{fig:rgb} and \ref{fig:cont}),
which continuum emission is unlikely dominated by dust emission 
(See discussions below in Section \ref{sec:nonthfil}).

We used the $\sim$7\arcsec resolution NH$_{3}$ rotational 
temperatures ($T_{\rm NH_3}$) derived in paper~I. 
For the regions where the NH$_{3}$ data are not available, 
we assume a gas kinematic temperature of 
$\langle T_{\rm NH_3} \rangle$ = 15 K, the average gas 
temperature derived from the observed NH$_{3}$ data.

\subsection{Velocity Structures} 
\label{vstruct}
\begin{figure*}[ht!]
\epsscale{1.2}
\plotone{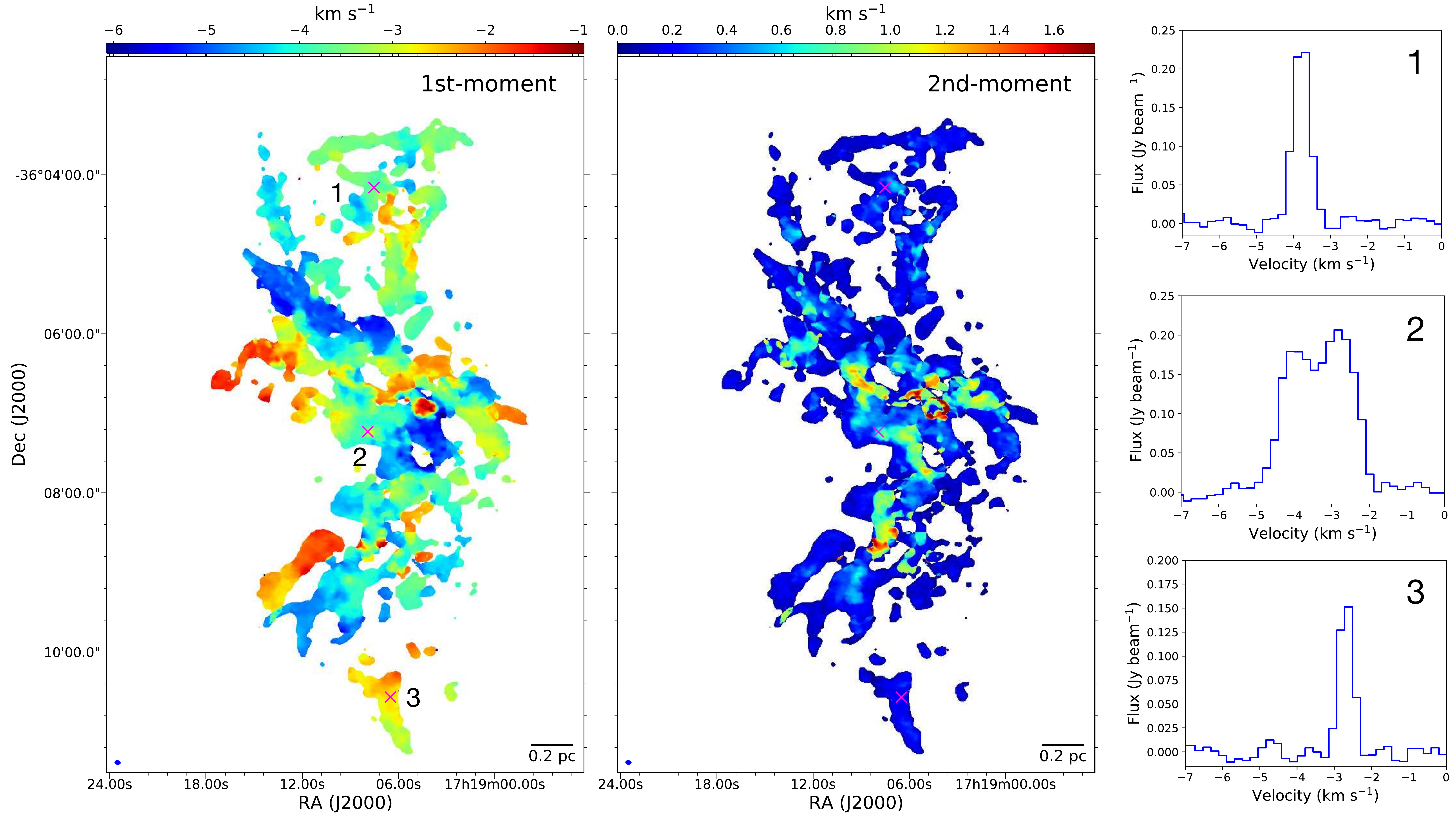}
\caption{Left and middle panels show the H$^{13}$CO$^{+}$  
intensity-weighted velocity (1st-moment) and 
intensity-weighted dispersion (2nd-moment) maps, respectively.  
The beam size is shown in the bottom left of each panel. 
Right panel shows the spectra of H$^{13}$CO$^{+}$  extracted 
from positions 1, 2, and 3. Three selected positions are marked 
with red cross ``x" in both left and middle panels. 
\label{fig:mom1}}
\end{figure*}

Paper~I found that multiple velocity components were detected in 
some areas where significant H$^{13}$CO$^{+}$ emission was 
detected.
Since the majority ($\sim$85\%) of H$^{13}$CO$^{+}$ emission 
appears as a 
single velocity component, we show the intensity-weighted velocity 
(1st-moment) and intensity-weighted dispersion (2nd-moment) of 
the H$^{13}$CO$^{+}$ line emission in Figure~\ref{fig:mom1}.  
We note that there are complex velocity structures across 
NGC\,6334S, especially toward the central region which appears
to be associated with multiple velocity components.

We fit Gaussian line profiles to the H$^{13}$CO$^{+}$ data 
pixel by pixel with multiple velocity components, under the 
assumption that the H$^{13}$CO$^{+}$ emission is optically thin. 
The detailed fitting process of molecular 
lines is summarized in the Paper~I. 
The observed velocity dispersions ($\sigma_{\rm obs}$) derived 
from the Gaussian fitting are between 0.10 and 0.80 \kms, with 
mean and median values of 0.23 and 0.20 \kms, respectively.  
The observed $\sigma_{\rm obs}$ is composed of the thermal 
and nonthermal components. 
Paper I shows that the nonthermal velocity dispersion 
$\sigma_{\rm nth}$ is dominated by 
subsonic and transonic motions throughout NGC\,6334S. 
The $\sigma_{\rm obs}$ of the dense cores is greater 
than in the quiescent regions;  the $\sigma_{\rm nth}$ 
and the $\sigma_{\rm obs}$ toward the central 
region of NGC\,6334S, have generally larger values than 
that measured in the outer regions (see Paper~I).

\subsection{Filament Identification} 
\label{sec:filident}
\subsubsection{Friend-of-Friend Algorithm} 
\label{sec:fof}
%
The results from the Gaussian fitting as described above 
in Section~\ref{vstruct} were used to identify gas filaments.  
Following similar procedures as   
\cite{2013A&A...554A..55H,2018A&A...610A..77H}, 
we used the python-based friend-of-friend (fof) 
algorithm\footnote{\url{https://github.com/ShanghuoLi/pyfof}} 
\citep{1982ApJ...257..423H} 
to identify the velocity-coherent filaments, 
i.e., no abrupt change of sign of the gradient along the 
filament in PPV space.

We first used the fof algorithm to identify the seed 
points, those that have peak intensities ($I$, the maximum 
intensity of the spectrum) above a certain threshold $I_{0}$ 
(7$\sigma$), of each individual structure. 
In total, about $\sim$70\% of the data points are above $I_{0}$.  
The spatial criterion between nearby points to be considered 
as friends is $\Delta r  \leqslant $ 0.023 pc ($\sim$1 beam 
size linear scale), 
while the velocity criterion uses an adaptive velocity gradient 
$\bigtriangledown v_{{\rm LSR},i} = 
\frac{1}{2} \frac{\bigtriangleup v_{i}}{\theta_{\rm FWHM}}$ 
similar to the definition in \cite{2018A&A...610A..77H}. 
Here, $\bigtriangleup v_{i}$ is the line full width half 
maximum (FWHM) of H$^{13}$CO$^{+}$ of the $i$th pixel and 
$\theta_{\rm FWHM}$ is the beam size. Only structures that 
contain more than 150 data points (the area of a structure 
larger than 3 times  the beam size) were considered. 
Second, we ran fof again to search for new friend points of each 
group identified in the first step, where the new friend points 
come from the remaining data points, in which the low intensity 
points  ($\leqslant I_{0}$) are encompassed; the same spatial and 
velocity criteria are used.  
After the second fof run, there are about 20\% of points that 
are not included within any group. The majority of them have 
relatively low intensities and/or appear to unaffiliated with  
the identified filaments.

We employed the 
\texttt{FilFinder}\footnote{\url{https://github.com/e-koch/FilFinder}} 
algorithm to compute the filament spine. The \texttt{FilFinder} 
package reduces the masking area to identify a skeleton that 
represents the topology of the area, using a Medial Axis 
Transform.  The masking area is delineated by the spatial 
distribution of the identified filaments;  we refer to the 
derived skeletons as the filament spines. The derived filament 
spines are shown  in Figure~\ref{fig:rgb}. 
In total, 13 velocity-coherent filaments have been identified 
by the fof algorithm from the PPV space of H$^{13}$CO$^{+}$ 
data.  Filaments are named F1, F2, F3 ... in order from south 
to north.  Filaments F4 and F10 have additional branches and 
they are named as F4b and F10b.  The physical 
lengths of these identified filaments ($L_{\rm fil}$) 
range between 0.4 and 1.3 pc.

\begin{figure*}[ht!]
\epsscale{1.15}
\plotone{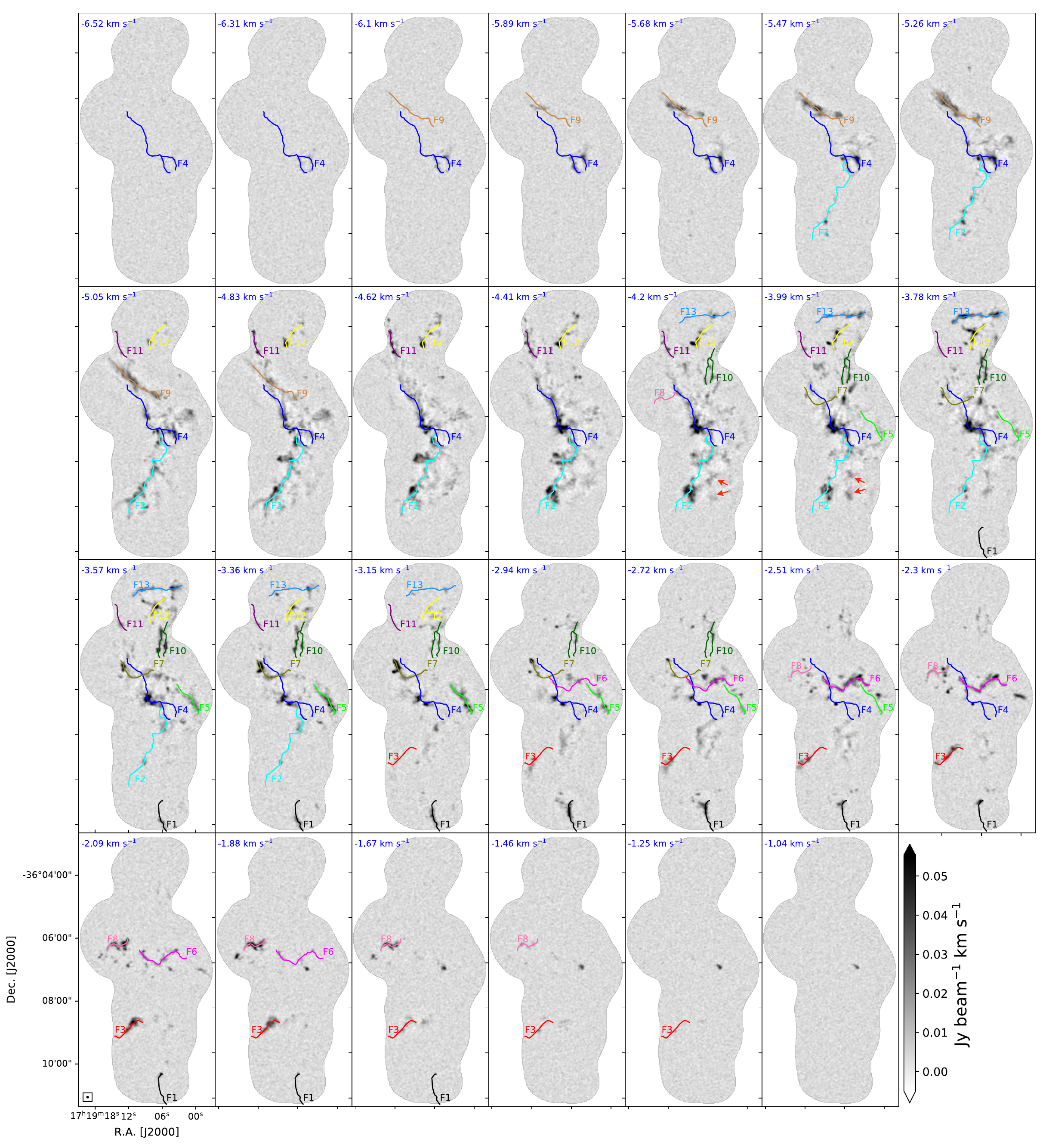}
\caption{The filament spines (solid curves) overlaid on the 
channel map of H$^{13}$CO$^{+}$. The velocity value is 
presented in the top left of each panel. 
\label{fig:chanmap}}
\end{figure*}

\subsubsection{Velocity-Coherent Filaments} 
\label{sec:vcoh}
\begin{figure*}[ht!]
\epsscale{1.}
\plotone{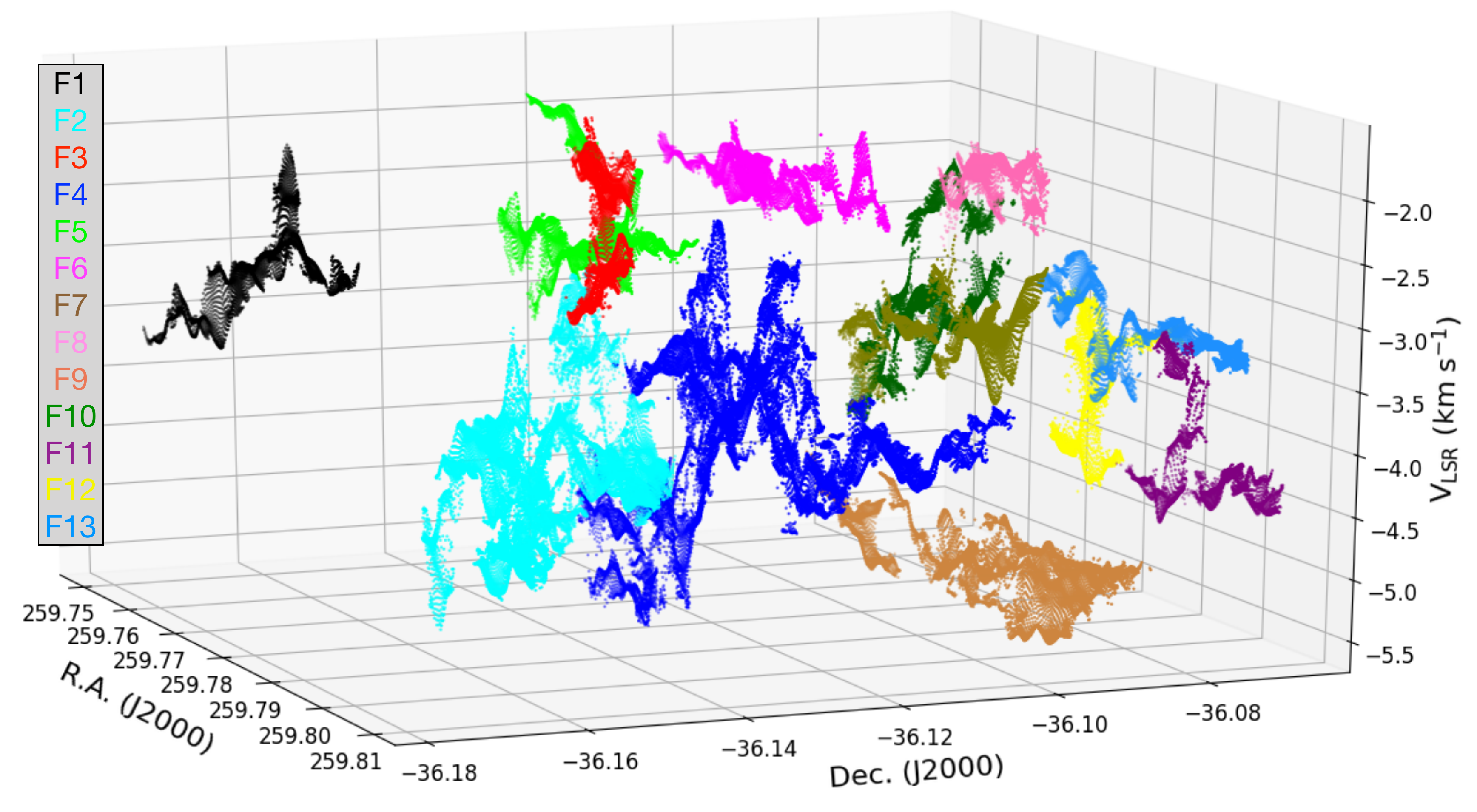}
\caption{Position-position-velocity (PPV) 
cube shows the centroid velocity of the identified 
filaments. (The animated version of the PPV cube is 
available in 
\url{https://github.com/ShanghuoLi/NGC6334S-filament}.)
\label{fig:fil3d}}
\end{figure*}

The filament spines overlaid on the channel maps of 
H$^{13}$CO$^{+}$ are shown in Figure~\ref{fig:chanmap},  
which shows that the line emission 
exhibits a filamentary distribution and that the identified 
filaments are consistent with the majority of the 
H$^{13}$CO$^{+}$ emission. This provides further evidence 
that the python-based fof algorithm can accurately  
recover the gas filamentary structures that are 
connected in both velocity and space (see also 
Figure~\ref{fig:fil3d}).

Several small regions show significant 
H$^{13}$CO$^{+}$ emission but are not grouped into any 
identified filament.  For instance, there is a small H$^{13}$CO$^{+}$ 
emission region on the west of F2 at velocity range of between 
-4.20 and 3.78~\kms, which is marked with red arrows in  
Figure~\ref{fig:chanmap}.  Two separated substructures appear 
in this small region, implying the emission is not connected in  
the spatial space.  
There are also some isolated small  regions separated 
from the identified filaments in  spatial space. These isolated 
regions fail to be considered as an independent filament  
because their emission is too weak 
and/or the number of total data points are lower than the 
criteria of identification.  We stress that the identified 
filaments are likely to be incomplete.  Potential low 
density and diffuse molecular filamentary structures could 
have been missed if their H$^{13}$CO$^{+}$ line emission 
is not significant and/or the detection suffers from 
severe missing flux.

The H$^{13}$CO$^{+}$ line emission of identified filaments  
generally spans $\geqslant$~4 channels  
(a velocity range of $\geqslant$~0.84~\kms).  
The filament with the largest spread in velocity is F4, which 
spans from -6.52 to -2.3 \kms.  
The majority of the filaments  are spatially distinct, 
while several filaments partly overlap in 
position, such as F2--F4, F4--F6, F4--F7, and F7-F8. 
The overlapping regions tend to show complex velocity structures 
as characterized by multiple velocity components along the line 
of sight.

\begin{figure*}[ht!]
\epsscale{1.1}
\plotone{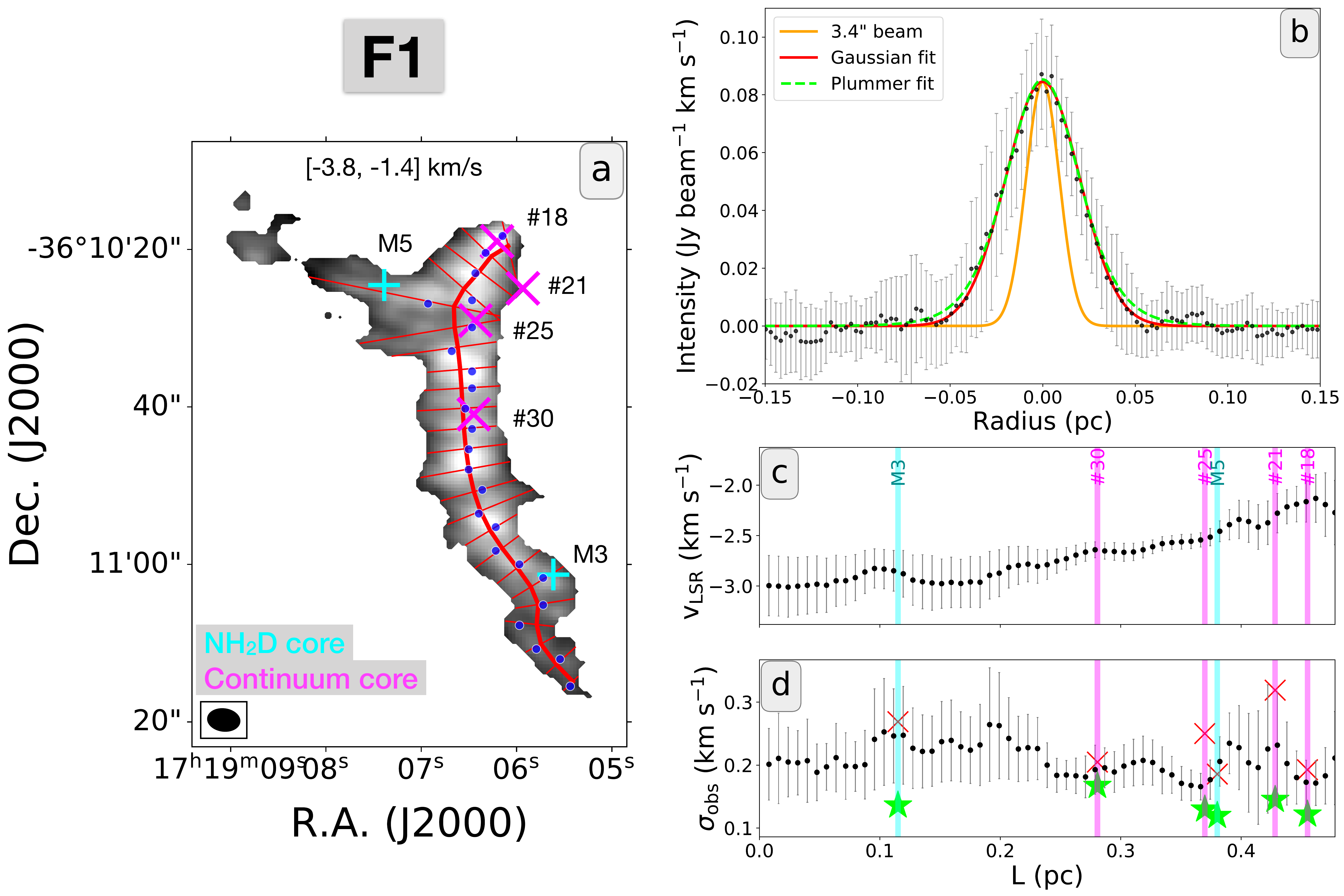}
\caption{Panel a: the filament spine (red solid curves) of F1 
overlaid on the  velocity-integrated intensity map of 
H$^{13}$CO$^{+}$.  Magenta cross ``x" and cyan plus ``+" 
symbols are continuum cores and NH$_{2}$D cores, respectively.  
The velocity range of this filament is presented in the upper 
middle of the panel. 
The beam size is shown in the bottom left of the panel. 
Panel b: mean integrated intensity profile of H$^{13}$CO$^{+}$ 
(black dots) was built by sampling radial cuts 
(short red solid lines) every 8 pixels (which corresponds to 
3\arcsec.44 or $\sim$0.019 pc at the source distance of 
1.3 kpc) along the spine.  The radial distance is the 
projected distance from the peak emission at a given cut 
(blue dots in panel a). The error bar represents 
the standard deviation of the cuts at each radial distance. 
The orange solid line shows the beam response with a 
FWHM of $\sim$3\arcsec.4. 
The red solid and green dashed lines present the best-fit 
results of Gaussian and Plummer fitting, respectively. 
Panel c and d: the mean centroid velocity 
$\langle v_{\rm LSR} \rangle$ and mean observed 
velocity dispersion $\langle \sigma_{\rm obs} \rangle$  of 
H$^{13}$CO$^{+}$ line variation along the filament. 
The error bars show the standard deviation of corresponding 
$v_{\rm LSR}$ and $\sigma_{\rm obs}$. 
Vertical magenta and cyan lines indicate the positions of 
associated continuum cores and NH$_{2}$D cores, respectively. 
The red cross ``x"  and green filled star symbols mark the 
core mean $\sigma_{\rm obs}$ derived from the H$^{13}$CO$^{+}$ 
and NH$_{2}$D lines, respectively. 
The complete figure set (4 images, see Figure~\ref{fig:comb}) 
is available in the online journal. 
\label{fig:F1}}
\end{figure*}

\begin{figure*}[ht!]
\epsscale{1.2}
\plotone{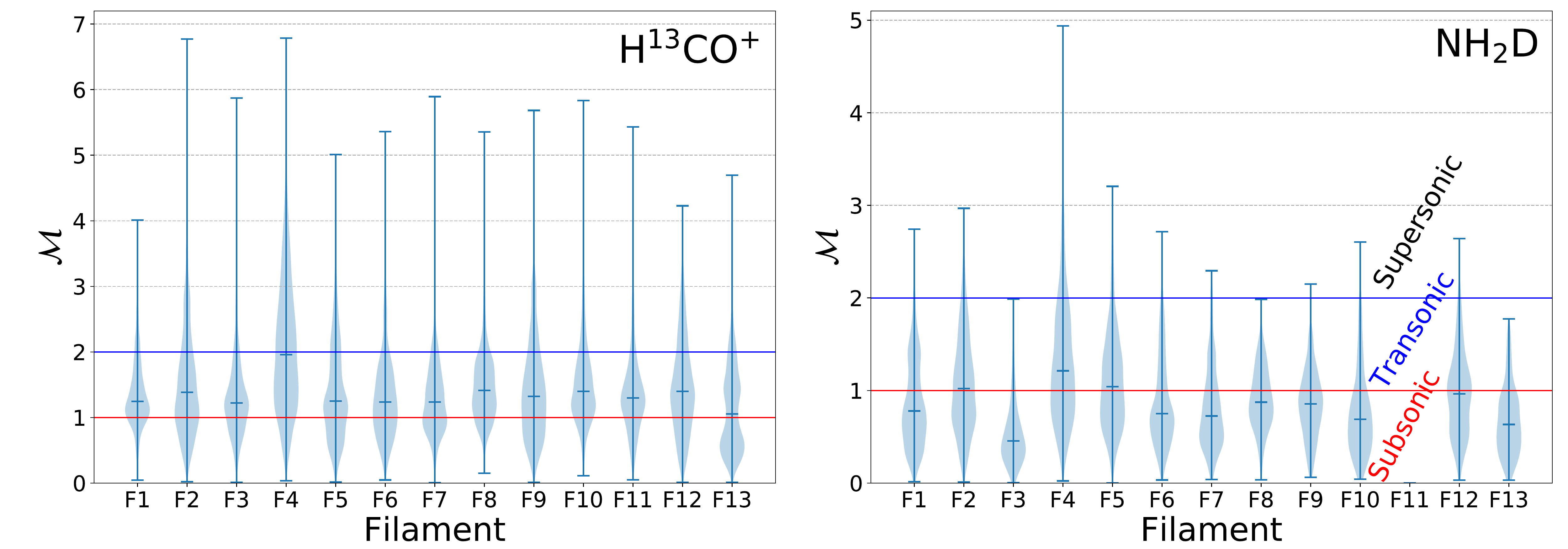}
\caption{Left and right: violin plots of the Mach number 
distributions derived from H$^{13}$CO$^{+}$ and NH$_{2}$D 
for each filament. The blue bars from the top to bottom 
represent the maximum, mean, and minimum values, 
respectively. The red and blue solid lines are the Mach 
number of 1 and 2, respectively. 
\label{fig:sig}}
\end{figure*}

\subsection{Filament Profile} 
\label{filprof}
We employed the 
\texttt{RadFil}\footnote{\url{https://github.com/catherinezucker/radfil}} 
package \citep{2018ApJ...864..152Z}, a radial density 
profile building and fitting tool for interstellar filaments, 
to construct the filament radial profile from  the velocity 
integrated intensity of H$^{13}$CO$^{+}$ inside the mask of a 
given filament.  
We used the \texttt{RadFil} tangent to the filament spine at 
7 or 8 pixel (about 1 beam size; 1 pixel = 0\arcsec.43) 
intervals along the filament, then took the radial cut perpendicular 
to each tangent. The radial profile has been shifted along each cut 
in order to ensure that it is centered on the pixel with the peak 
intensity. Figure~\ref{fig:F1} shows the radial cut and the pixels 
(blue points) of the peak in the radial cut for F1.  Along each cut, 
the radial distance is calculated as the projected distance from 
the peak intensity. Prior to fitting the profile, the background 
was subtracted using the background subtraction estimator of 
\texttt{RadFil}. 
The background is estimated by a first-order polynomial to all 
profiles at the given radial distance range, and then subtracts 
it from each cut; the background subtraction radii vary slightly 
from filament to filament, with a typical range of 0.08--0.15 pc.

To compute the filament widths (FWHM), we performed a  Gaussian 
fitting to the average profile of the H$^{13}$CO$^{+}$ intensity 
of each filament. The Gaussian function is given by 
\begin{equation}
\label{gauf}
A(r) = A_{0}\; \rm exp\left(\frac{-(r-\mu)^{2}}{2\sigma_{G}^{2}} \right),
\end{equation}
where $r$ is the radial distance, $A$ is the profile amplitude at 
the radial distance $r$, $A_{0}$ is the amplitude, $\sigma_{G}$ 
is the standard deviation,  and $\mu$ is the mean. Here, $\mu$ is 
fixed to zero. The best-fit Gaussian of each filament profile is 
listed in Table~\ref{tab:fil}.  An example of the fit is shown 
in Figure~\ref{fig:F1}, where the red solid line is the best fit, 
the black dots correspond to the averaged integrated intensity 
of H$^{13}$CO$^{+}$ and the error bars are the standard deviation 
of the radial profile of all cuts perpendicular to the filament.  
The best-fit filament widths range from 
0.036 to 0.074 pc, with the mean and median values of 0.046 and 
0.045 pc, respectively. 
We also estimated the beam-deconvolved FWHM with 
FWHM$_{\rm decon}$ = $\sqrt{\rm FWHM^{2}-FWHM_{\rm bm}^{2}}$, 
where FWHM$_{\rm bm}$ is the half-power beam width. 
The FWHM$_{\rm bm}$ is about 3\arcsec .4 ($\sim$0.021 pc) in 
our observations.  The FWHM$_{\rm decon}$ is  between 0.029 
and 0.071 pc, with the mean and median values of 0.041 and 
0.039 pc, respectively.

In previous studies, observed filaments have been  considered 
as  cylindrical structures that can be described by a 
Plummer-like function of the form 
\citep[e.g.,][]{2008MNRAS.384..755N,2011A&A...529L...6A,
2013A&A...550A..38P,2014MNRAS.445.2900S,2018MNRAS.478.2119L}: 
\begin{equation}
\label{gauf1}
\int fdv(r) = \frac{A_{0}}{ \left[1 + \left( \frac{r}{R_{\rm flat}} \right)^{2} \right]^{\frac{p-1}{2}} },
\end{equation}
where $\int fdv$ is the integrated intensity, $A_{0}$ is the peak 
profile amplitude, $R_{\rm flat}$ is the flattening radius, and $p$ 
is the power-law index of the density profile at large radii 
\citep{2016A&A...590A.110C,2018ApJ...864..152Z}.  
We also performed a Plummer fitting to the identified filaments 
in NGC 6334S. 
The best Plummer fit is shown in dashed green line in 
Figure~\ref{fig:F1}. The filament widths derived by the Plummer 
fitting are similar to those of the Gaussian fitting, and FWHM 
ranges from 0.03 to 0.066 pc, with the mean and median values of 
0.045 and 0.042 pc, respectively.  The FWHM$_{\rm decon}$ is 
between 0.023 and 0.062 pc, with the mean and median values of 
0.039 and 0.037 pc (Table~\ref{tab:fil}), respectively. 
$R_{\rm flat}$ is between 0.012 and 0.081 pc, with a mean and 
median values of 0.033 and 0.027 pc, respectively. 
Figure~\ref{fig:F1} shows that some of the filament profiles 
have relatively large dispersions due to non-uniform line 
intensities throughout the filaments.
The significant variations 
of H$^{13}$CO$^{+}$ emission across the filaments results in 
a poor fit in their profiles, e.g., F4 and F9.

The derived filament widths are similar to those of Musca 
\citep[$\sim$0.07 pc;][]{2016A&A...586A..27K}, 
Aquila/Polaris \citep[$\sim$0.04 pc;][]{2010A&A...518L.103M}, 
Orion \citep[$\sim$0.02 -- 0.05 pc for OMC-1/2 and 
ISF;][]{2018A&A...610A..77H}, 
G14.225-0.506 \citep[$\sim$0.05 -- 0.09 pc;][]{2019ApJ...875...24C}, 
G035.39-00.33 \citep[$\sim$0.028 pc;][]{2017MNRAS.464L..31H}, 
and L1287 \citep[$\sim$0.03 pc;][]{2020A&A...644A.128S}. 
In contrast, the derived filament widths are narrower than 
those of \textit{Herschel} filaments studied 
toward IC\,5146 
\citep[$\sim$0.1 pc;][]{2019A&A...621A..42A}, Taurus 
\cite[$\sim$0.1 pc;][]{2013A&A...550A..38P}, 
NGC\,6334IN and NGC\,6334I 
\cite[$\sim$0.24 pc;][]{2013A&A...554A..42R}. 
But note that the spatial resolution of \textit{Herschel} 
observations (beam size $\sim$ 36\arcsec)  is much poorer than 
that of the ALMA observations. 
This supports the idea that higher spatial resolution observations and 
higher-density gas tracers can identify narrower filaments.  
In addition, the dust continuum emission cannot be resolved into 
separate filaments when they overlap, whereas velocities measurements 
generally can do so.  
Spatially blended filaments might broaden the measured filament 
widths \citep[see also][]{2017MNRAS.464L..31H}. 
This also indicates that filaments identified with different 
procedures might show deviating filament widths.

\subsection{Filament Mass} 
\label{sec:fmass}
The mass per unit length is one important indicator for assessing 
the stability of filaments.  The continuum emission from dust 
is one of the most frequently used measurement to compute the 
mass. Figure~\ref{fig:cont} shows the filament spines overlaid on 
the 3~mm continuum image.  Unfortunately, only 2 gas filaments (F4 
and F1) have significant continuum emission detection 
(Figure~\ref{fig:cont}); the remaining 11 gas filaments are 
either only partly detected in continuum emission or not 
detected at all above $5\sigma$.   An alternative 
way to estimate the filament mass is to make use of molecular gas 
emission; in this work we use H$^{13}$CO$^{+}$. With the 
fractional abundance of H$^{13}$CO$^{+}$ relative to H$_{2}$,  
$X(\rm H^{13}CO^{+})$ = $N(\rm H^{13}CO^{+})$/$N(\rm H_{2})$,   
the filament mass can be computed as follows:  
\begin{equation}
\label{Mfil}
M_{\rm fil} = \mu_{\rm H_2} \, m_{\rm H}  \sum \frac{ N(\rm H^{13}CO^{+}) }{X(\rm H^{13}CO^{+})} \, \Omega,
\end{equation}
where  $N(\rm H^{13}CO^{+})$ is the column density of 
H$^{13}$CO$^{+}$, $\mu_{\rm H_2}$ = 2.8 is the mean molecular 
weight of the interstellar medium \citep[ISM;][]{2008A&A...487..993K}, 
$m_{\rm H}$ is the hydrogen mass, and $\Omega$ is the solid 
angle of the H$^{13}$CO$^{+}$ emission.   Assuming local 
thermodynamic equilibrium (LTE), the molecular column 
densities can be estimated from the velocity-integrated intensity 
(see Appendix~\ref{column}), and finally leads to $M_{\rm fil}$ 
(see Eq.\ref{Mfil}).

In order to estimate $X(\rm H^{13}CO^{+})$ for NGC\,6334S, 
we have focused on F4 because both H$^{13}$CO$^{+}$ line and 
continuum emission are significantly detected 
(Figures~\ref{fig:rgb} and \ref{fig:cont}). 
The derived  $N(\rm H^{13}CO^{+})$ ranges from 
7.2$\times 10^{11}$~cm$^{-2}$ to 1.4$\times 10^{13}$~cm$^{-2}$ 
and $N_{\rm H_{2}}$ is between 1.3$\times 10^{22}$~cm$^{-2}$ and 
6.1$\times 10^{23}$~cm$^{-2}$. The resulting values of 
$X(\rm H^{13}CO^{+})$ extend from 7.7$\times 10^{-12}$ to 
3.1$\times 10^{-10}$, with a median value of 5.4$\times 10^{-11}$. 
The derived $X(\rm H^{13}CO^{+})$ is similar to the reported 
values of 3.0$\times 10^{-11}$ -- 4.0$\times 10^{-10}$ in 
\cite{1995ApJ...448..207B}, 
4.5$\times 10^{-11}$ in \cite{2014A&A...563A..97G}, 
4.8$\times 10^{-10}$ in \cite{2012ApJ...756...60S}, 
and 1.3$\times 10^{-10}$ in \cite{2013ApJ...777..157H}.

Using the median $X(\rm H^{13}CO^{+})$ = 5.4$\times 10^{-11}$, 
we estimated the gas mass ($M_{\rm fil}$) for each filament. 
The derived masses are in the range of 4 -- 82 \Mo\ 
(see Table~\ref{tab:fil})  and the total gas mass in the filaments 
is about 342~\Mo.  The total gas mass estimated from 
H$^{13}$CO$^{+}$ in the observed region is about 395 \Mo, 
which indicates that these filaments contain most of the dense 
gas (87\% = 342/395), as revealed by the total H$^{13}$CO$^{+}$ 
line emission (see Section 4.5 below).  
The masses per unit length 
($M_{\rm line} = M_{\rm fil}/L_{\rm fil}$) of filaments range  
between 14 and 64 \Mo~pc$^{-1}$,  with a median value of 
29 \Mo~pc$^{-1}$ (see Table~\ref{tab:fil}).

The uncertainties in the distance, assumed gas temperature, and 
variations of the  H$^{13}$CO$^{+}$ fractional abundance   
introduce uncertainties in the estimates of the filament masses 
and masses per unit length. The typical uncertainties in the 
temperatures derived from NH$_{3}$ is  $\sim$15\% (see 
Paper~I).  The uncertainty in distance from the trigonometric 
parallax measurement is $\sim$20\% \citep{2014ApJ...784..114C}. 
The standard deviation (std) of $X(\rm H^{13}CO^{+})$ for F4 
is about $3.7\, \times \, 10^{-11}$, which corresponds to 
1$\sigma$ uncertainty of 
70\% = $\frac{3.7\, \times \, 10^{-11}}{5.4\, \times \, 10^{-11}}$. 
We settle on an uncertainty estimate of a factor of $\sim$2 for 
both filament mass and mass per unit length according to 
the propagation of error. 
Considering of the inclination angle is unknown, the 
uncertainties in the mass per unit length could be larger.

\subsection{Subsonic and Transonic Filaments} 
\label{sonic}
The three-dimensional (3D) Mach number is 
$\mathcal{M} 
= \sqrt{3} \sigma_{\rm nth}/c_{\rm s}$, where $\sigma_{\rm nth} = 
\sqrt{\sigma_{\rm obs}^2 - (\triangle_{\rm ch}/2\sqrt{\rm 2ln2})^2 - 
\sigma_{\rm th}^2}$ is the nonthermal velocity dispersion,  
$c_{\rm s}$ is the sound speed, and $\triangle_{\rm ch}$ is the 
channel width.  The molecular thermal velocity dispersion can 
be estimated by 
$\sigma_{\rm th} = \sqrt{(k_{\rm B}T)/(\mu m_{\rm H})} = 
0.098\, {\rm km\,s^{-1}} \, (\frac{T}{\rm K})^{0.5} \mu^{-0.5}$, 
where $\mu = m/m_{\rm H}$ is the molecular weight, $m$ is the 
molecular mass, $m_{\rm H}$ is the hydrogen mass, and $T$ is 
the gas temperature (see also Paper~I). 
The sound speed $c_{\rm s}$ was estimated using a mean 
molecular weight per free particle of 
$\mu_{\rm p}$=2.37 \citep{2008A&A...487..993K}. 
Figure~\ref{fig:sig} shows the Mach number 
($\mathcal{M}$) distributions derived from H$^{13}$CO$^{+}$ and 
NH$_{2}$D for each filament, except for F11 which has no significant 
NH$_{2}$D detection. Some filaments  are 
partially overlapping but the corresponding NH$_{2}$D line emission 
shows only one  velocity component clearly.  
This NH$_{2}$D emission is assigned to a particular filament based 
on the minimum velocity differences between NH$_{2}$D emission 
and that filament. The Mach number distributions 
derived from both H$^{13}$CO$^{+}$ and NH$_{2}$D lines reveal 
that the majority of filaments are subsonic 
($\mathcal{M} \leqslant 1$) and transonic 
($1 < \mathcal{M} \leqslant 2$) in nonthermal motions.   
In general, the $\mathcal{M}$ derived from 
NH$_{2}$D tends to be smaller than those from H$^{13}$CO$^{+}$.  
This is because NH$_{2}$D emission traces colder, denser gas and it 
is less affected by the protostellar feedback (e.g., outflows) as 
compared to the H$^{13}$CO$^{+}$ emission; this is confirmed 
by the fact that  the observed line widths of the former 
narrower than those of the latter.  The subsonic and transonic 
features imply a quiescent nature of these filaments.  
The subsonic and transonic nonthermal line widths found here 
in dense filaments and dense cores have been seen previously 
in low-mass star-forming 
regions (e.g., Perseus, \citealt{2010ApJ...712L.116P}; 
Serpens, \citealt{2021A&A...646A.170G}; Ophiuchus and Taurus, 
\citealt{2019ApJ...877...93C}; 
L1478 in the California, \citealt{2019ApJ...877..114C}),  
intermediate- and high-mass 
star-forming regions (e.g., Orion, 
\citealt{2018A&A...610A..77H,
2018ApJ...861...77M,2021RAA....21...24Y};  
IRDC G035.39-00.33, \citealt{2018A&A...611L...3S};  
IRDC G14.225–0.506, \citealt{2019ApJ...875...24C}).

\begin{figure}[ht!]
\epsscale{1.2}
\plotone{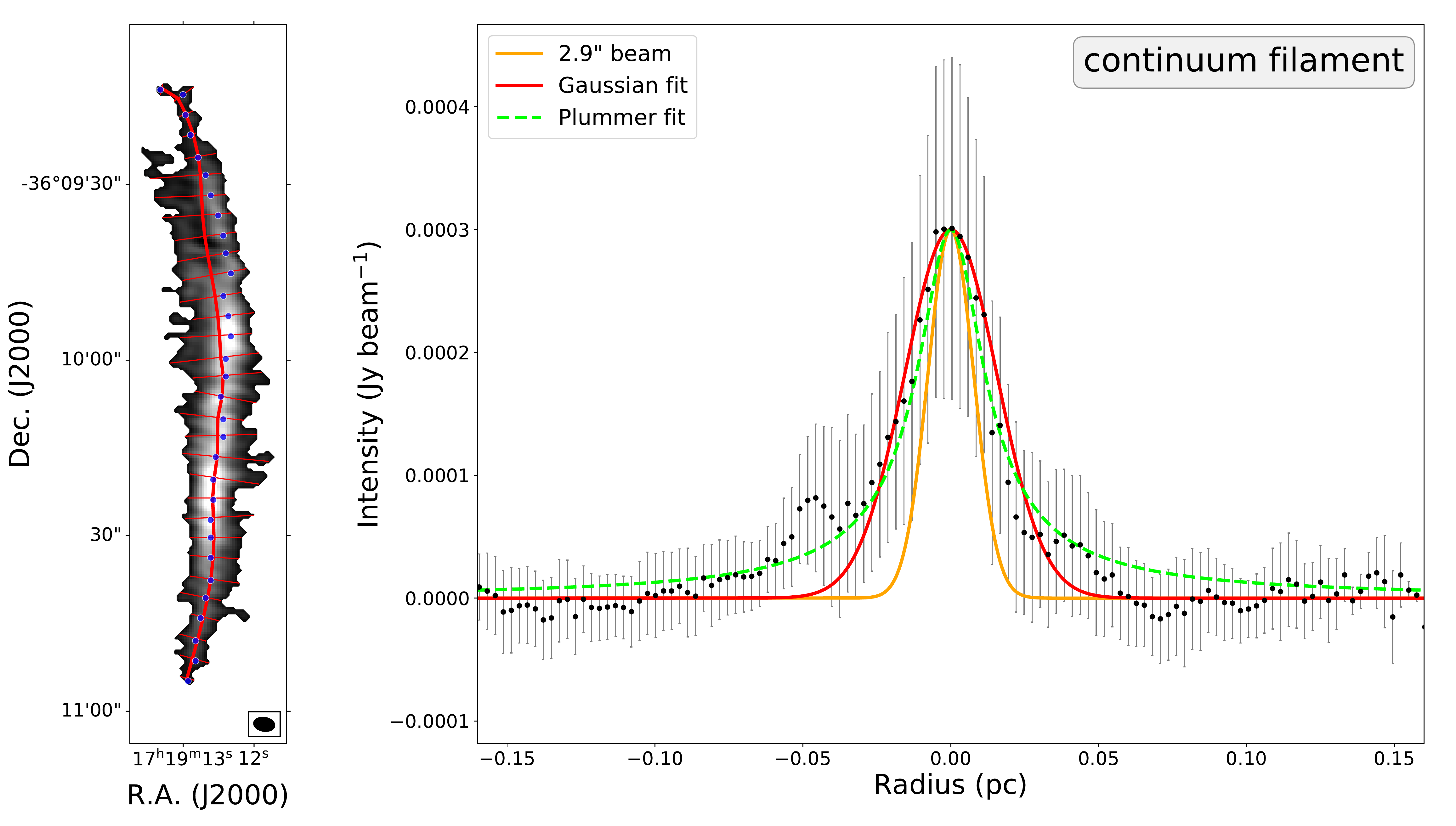}
\caption{Left: the filament spine (red solid curve) of continuum 
filament overlaid on the 3~mm continuum image. 
Right: mean integrated intensity profile (black dots) was built 
by sampling radial cuts (short red solid lines) every 8 pixels 
(3\arcsec.44 corresponds to $\sim$0.019 pc at the source 
distance of 1.3 kpc) along the spine.  The radial distance is 
the projected distance from the peak emission at a given cut 
(blue dots in the left panel). The error bar represents 
the standard deviation of the cuts at each radial distance. 
The orange solid line shows the beam response with a 
FWHM of $\sim$2.9\arcsec. 
The red solid and green dashed lines present the best-fit 
results of Gaussian and Plummer fitting, respectively. 
\label{fig:contfil}}
\end{figure}

\section{Discussion}
\label{sec:dis}

\subsection{Continuum Filament}
\label{sec:nonthfil}

The majority of the continuum emission structures have a significant 
line emission counterparts.  One exception is the continuum 
filamentary structure in the south-eastern part of the map 
(see Figures~\ref{fig:rgb} and \ref{fig:cont}), which 
has no line emission 
counterpart except for a small area near the middle  that 
shows weak emission in the H$^{13}$CO$^{+}$, CS, HCN and 
HCO$^{+}$ lines. If this filamentary structure's continuum 
emission was dominated by free-free or synchrotron emission 
rather than dust emission, the stellar feedback from O type stars 
\citep[O7.5 and O6.5, see 
Figure~\ref{fig:rgb};][]{2008hsf2.book..456P}  
on the north-eastern side of NGC\,6334S are most likely 
responsible.

We used the \texttt{FilFinder} to extract this filamentary 
structure and its filament spine (see Figure~\ref{fig:contfil}). 
The filament length is about 0.8 pc. 
Using \texttt{Radfil}, the filament widths are about 0.032 pc 
and 0.023 pc derived by Gaussian and Plummer fitting based 
on the continuum emission (Table~\ref{tab:fil}), respectively. 
The gas mass cannot be reliably estimated because of the 
unknown fraction of dust emission.

\subsection{The Kinematics of Filaments} 
\label{sec:kinfil}
Figure~\ref{fig:F1} shows the variations of $v_{\rm LSR}$ and 
$\sigma_{\rm obs}$ derived from the H$^{13}$CO$^{+}$ line 
emission along the filament.  Only one filament (F1) shows 
monotonic  changes along the filament in $v_{\rm LSR}$. 
In contrast,  the $\sigma_{\rm obs}$ shows small fluctuations 
along F1 rather than  monotonic  change. Four continuum cores and 
two  NH$_{2}$D cores are associated with this filament. The 
core mean $\sigma_{\rm obs}$ derived from the H$^{13}$CO$^{+}$ 
line is comparable to the $\sigma_{\rm obs}$ of filament, 
however  continuum cores \#25 and \#21  have slightly larger 
$\sigma_{\rm obs}$. Line widths may be broadened by active 
star formation. In contrast, some cores show much 
narrower $\sigma_{\rm obs}$ than their respective filament; 
e.g., \#31/\#36/\#49 in F2, \#4/\#24 in F4, M11 in F5, M10 in F6, 
M14 in F10.  The measured $\sigma_{\rm obs}$ of NH$_{2}$D 
is always narrower than those of H$^{13}$CO$^{+}$ toward the 
continuum cores and NH$_{2}$D cores, and this feature is also 
seen in the filaments (see Figure~\ref{fig:sig}). 
In Section~\ref{sec:kincore}, we will discuss the properties of
$\sigma_{\rm obs}$ in both continuum cores and NH$_{2}$D cores.

\begin{figure}[ht!]
\epsscale{1.1}
\plotone{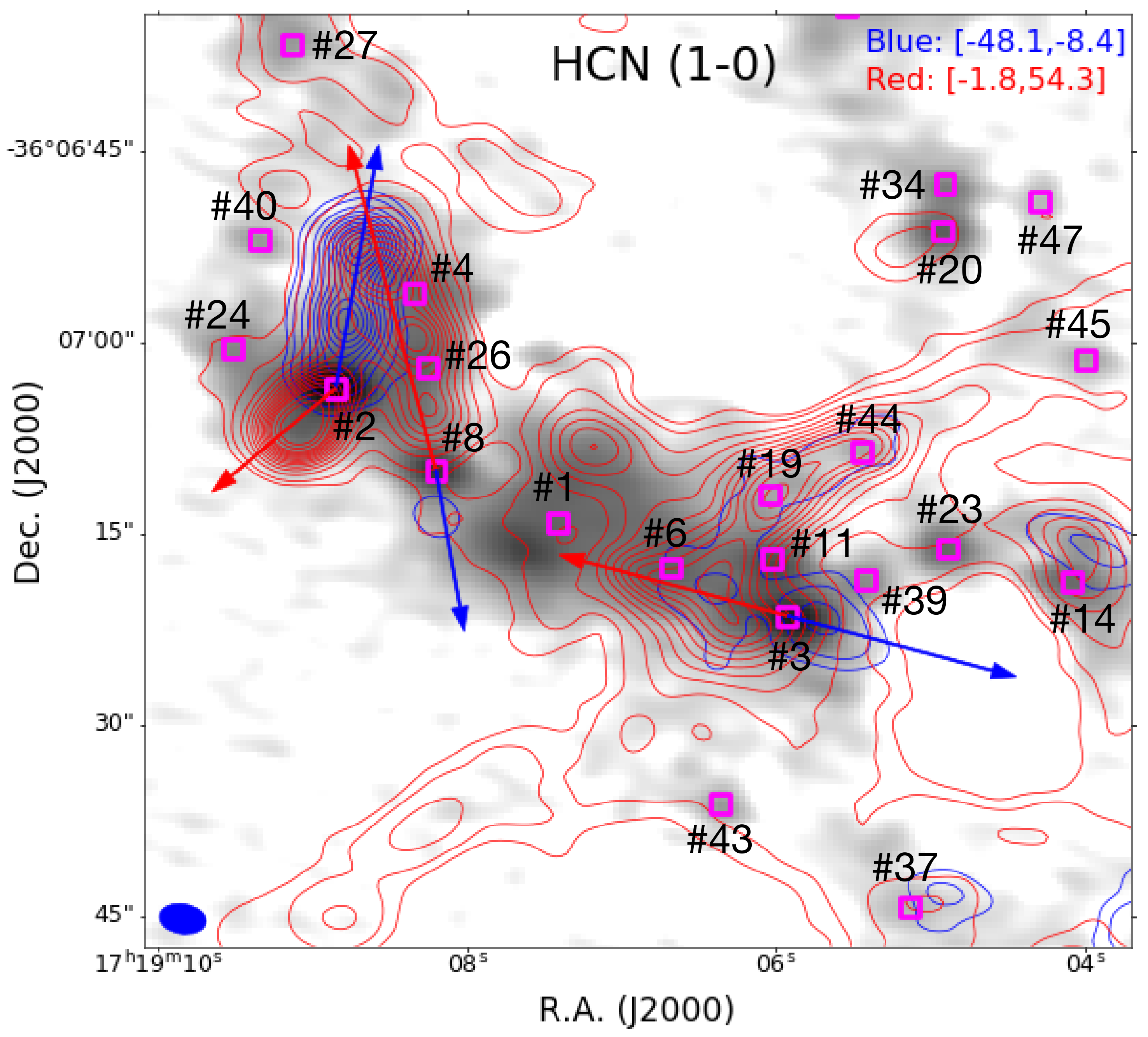}
\caption{Magenta open squares show the continuum cores. 
The blue-shifted (blue contours) and red-shifted (red 
contours) of HCN (1-0) line emission overlaid on the continuum 
image. The arrows show the outflow directions. 
The beam size is shown in the bottom left of each panel.
 \label{fig:outflow}}
 \end{figure}

Some filaments show only small $v_{\rm LSR}$ variations along 
their spine (F4b, F5, F7, F8, F10, F10b, F12), while others  
display significant variations (F2, F3, F4, F6, F9, F11, F13).  
The $\sigma_{\rm obs}$ also shows irregular variations in all of 
filaments, except for F1 and F13. The latter shows roughly an 
increasing trend from east to west (Figure~\ref{fig:comb}).
There is another gaseous structure seen in the CS (1-0), 
HCO$^{+}$ (1-0), and HCN (1-0) lines, which runs 
northwest to southeast through the western part of F13.  
The interaction between two gas flows can broaden the 
H$^{13}$CO$^{+}$ line widths around the intersection regions. 
On the other hand, the star formation activity is  responsible 
for some of the $v_{\rm LSR}$ and $\sigma_{\rm obs}$ variations 
in some filaments. For instance, F4 shows $v_{\rm LSR}$ and 
$\sigma_{\rm obs}$ variations at positions in which  several 
strong molecular outflows are detected in the HCN, HCO$^{+}$, 
and CS lines (see Figure~\ref{fig:outflow}). 
These molecular outflows can inject energy and momentum into 
the immediate surroundings of protostars and affect the gas 
kinematics, and then the turbulence (or line width) will be increased 
and the gas velocity will be modified \citep{2019ApJ...878...29L,
2020ApJ...903..119L,2021ApJ...909..177L}. Filament F4 
encompasses the highest number of continuum cores and YSOs 
among the filaments (Figures~\ref{fig:rgb} and \ref{fig:comb}), and 
therefore the $v_{\rm LSR}$ and $\sigma_{\rm obs}$ of the 
H$^{13}$CO$^{+}$ line emission in F4 is most likely  
significantly affected by protostellar feedback. 
Overall, both protostellar activity and interaction between 
gas flows can significantly alter the local gas kinematics.

\subsection{Velocity Gradient Along F1} 
\label{sec:vgrad}
As shown in Figure~\ref{fig:F1}, F1 presents a  smooth 
velocity change in H$^{13}$CO$^{+}$  emission from the south 
(-3.3 \kms) to the north (-1.8 \kms) along the filament, resulting 
in a projected velocity gradient of $\sim$1.8$\pm$0.1~\kms~pc$^{-1}$.  
The NH$_{2}$D line emission also shows a similar velocity gradient 
along F1.  The velocity gradient along F1 could be attributed 
to the ongoing accretion flow in F1 
\citep[e.g.,][]{2013ApJ...766..115K},  
whereas we cannot completely rule out the possibility that 
the gas kinematics is affected by the external feedback from the 
western YSOs, such as molecular outflows and/or expanding shells.

If the velocity gradient comes from the accretion flow along 
the F1 filament, one can estimate the mass flow rate using the 
derived velocity gradient and filament mass. 
Assuming that the filament has a cylindrical geometry, the mass 
flow rate, $\dot{M}$, can be calculated as 
\citep[see][]{2013ApJ...766..115K} 
\begin{equation}
\label{accretion}
\dot{M}\, = \, \frac{M \, \nabla_{||} v_{\rm obs} }{ {\rm tan}(\alpha) } 
\end{equation}
where $M$ is the filament mass, $\nabla_{||} v_{\rm obs}$ is the 
observed velocity gradient along the filament, and $\alpha$ is 
the angle of inclination to the plane of sky. Using the derived 
filament mass of 14 \Mo, the observed velocity gradient of 
1.8 \kms pc$^{-1}$, and assuming a moderate inclination angle 
of $\alpha$ = 45$^{\circ}$, the mass flow rate is estimated 
to be about  26 \Mo Myr$^{-1}$ for F1. This result indicates that 
the F1 filament will double its mass in several free-fall time; 
$\sim$1$\times$10$^{5}$ yrs assuming a density of 
10$^{5-6}$ cm$^{-3}$ that is the typical value of continuum 
cores in the F1 (see Paper~I).

Considering the uncertainties of the derived mass and 
inclination angle, the estimated flow accretion is roughly 
comparable to 
the values of 70$\pm$40 \Mo Myr$^{-1}$ in the 
IRDC\,G035.39--00.33 \citep{2014MNRAS.440.2860H},  
and 17 -- 72 \Mo Myr$^{-1}$ in the Monoceros R2 
\citep{2019A&A...629A..81T}, and 20 -- 130 \Mo Myr$^{-1}$ 
in the IRDC\,G14.225--0.506  \citep{2019ApJ...875...24C}.  
The estimated mass flow rate could be treated as a lower 
limit, because the H$^{13}$CO$^{+}$ only traces relatively 
high dense gas and F1 filament is only a small part of a 
much large filamentary structure seen in the infrared image 
(see Figure~\ref{fig:rgb}).

A  velocity gradient is also detected in sections of the other filaments 
(e.g., F7 and F13) and around some embedded cores 
(e.g., \#29 in the F3; see Figure~\ref{fig:comb}).  
Unfortunately, in these cases we cannot distinguish whether the 
velocity gradient is the result of gas flow or some other 
physical process, e.g., molecular outflow or rotation. 
Thus, we refrain from estimating the mass flow rate 
for other filaments.

\begin{figure*}[ht!]
\epsscale{1.15}
\plotone{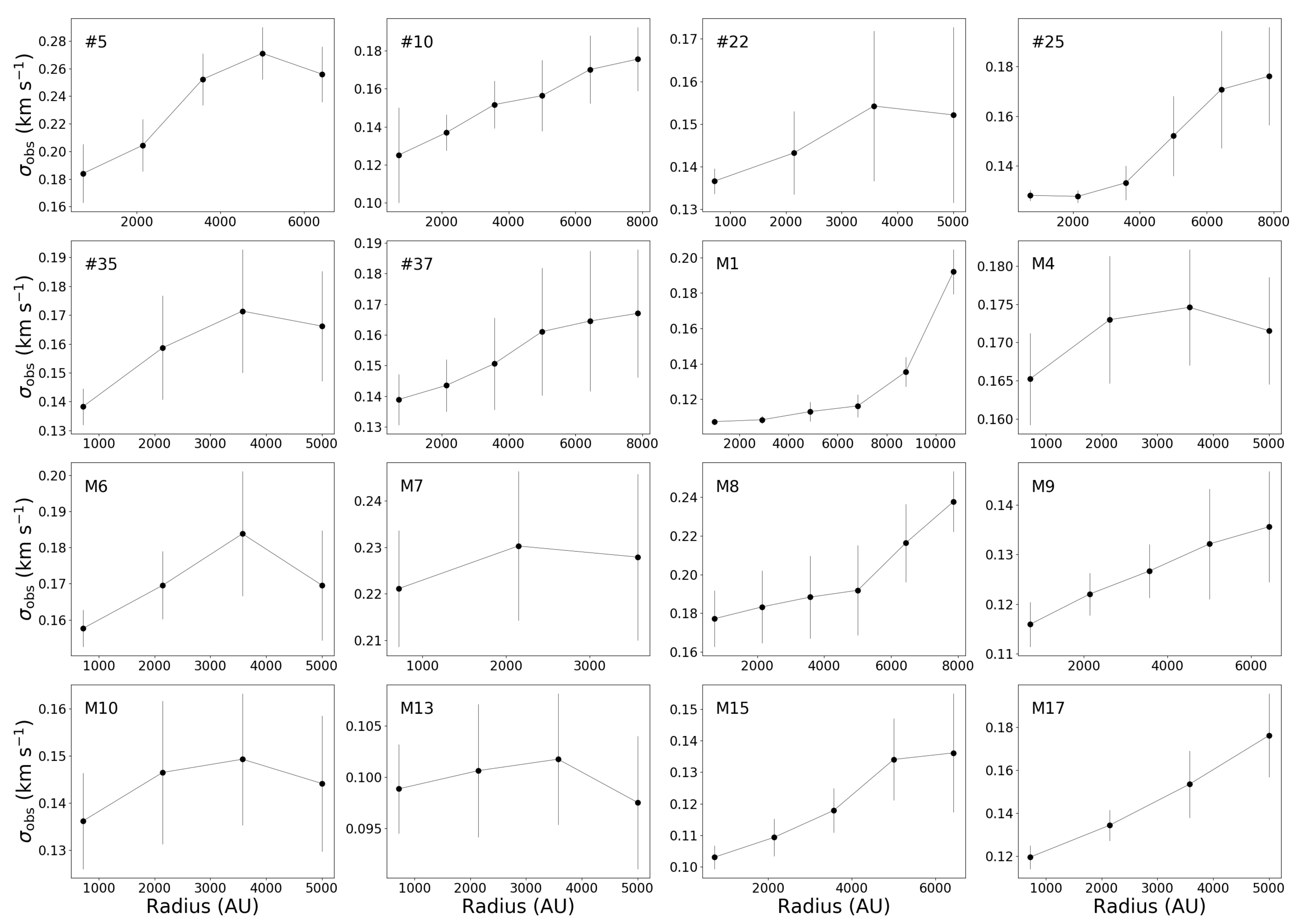}
\caption{The annularly averaged observed NH$_{2}$D velocity 
dispersions ($\sigma_{\rm obs}$) as a function of radial distance 
from the center of cores. The error bars show the statistical standard 
deviation inside each ring divided by the square root of the length 
of the ring. M1 is modified from Paper~II. 
The name of each core is shown in the top left of each panel. 
\label{fig:Rsig}}
\end{figure*}

\begin{figure*}[ht!]
\epsscale{1.2}
\plotone{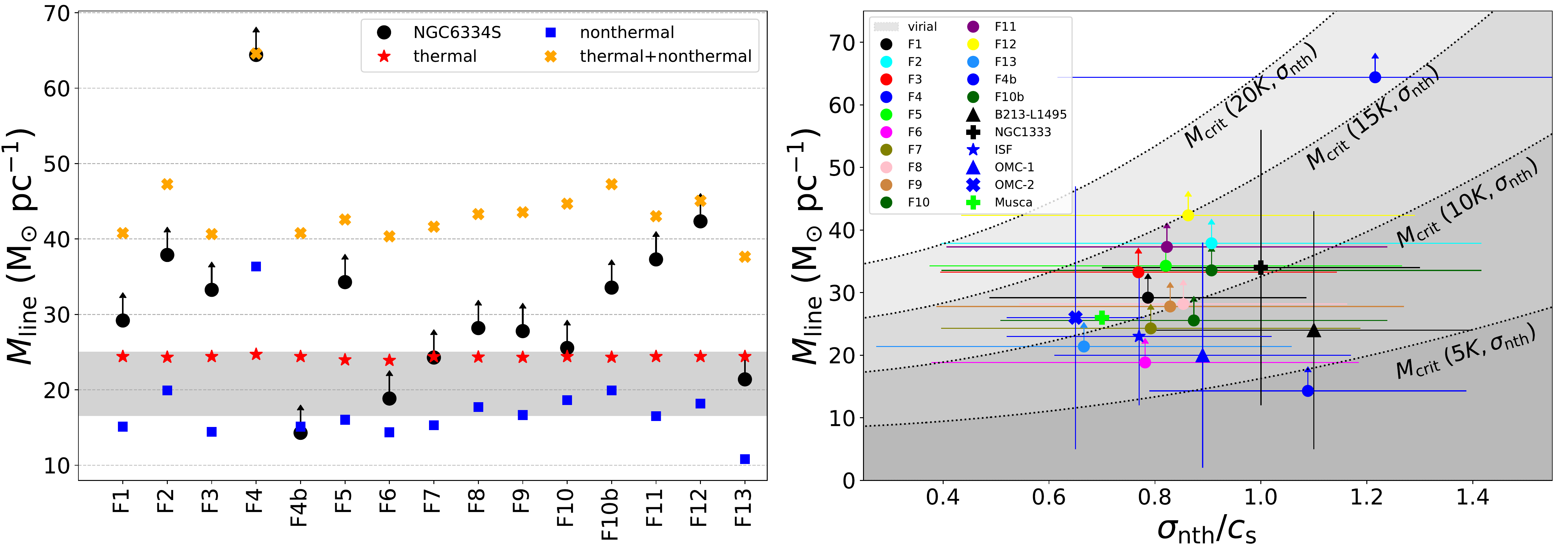}
\caption{
Left: the line-mass for each filament.  The black filled circles, red 
filled stars, blue filled squares, and orange filled crosses show the 
estimated mass per unit length, thermal critical line-mass, 
nonthermal critical line-mass,  and total (thermal + nonthermal) 
critical line-mass, respectively. 
The shaded gray region shows the thermal critical line-mass of 
16.6 -- 25 \Mo~pc$^{-1}$ corresponding to the gas temperature of 
10 -- 15 K. 
The arrows indicate that the estimated mass per unit length could  
be treated as a lower limit.
Right: mass per unit length vs. $\sigma_{\rm nth}/c_{\rm s}$. 
The error bars indicate the standard deviation of the parameters.  
The dashed lines show the expected total critical line-mass for an 
infinite filament in hydrostatic equilibrium at temperatures of 5 K, 
10 K, 15 K, and 20 K, respectively (see Appendix~\ref{ap:mcrit}). 
The data points of B213-L1495, NGC1333, ISF, OMC1, OMC2, 
and Musca are retrieved from \cite{2018A&A...610A..77H}. 
\label{fig:mass}}
\end{figure*}

\subsection{The Kinematics of Embedded Cores} 
\label{sec:kincore}
As mentioned in Section~\ref{sec:vgrad},  some cores 
show smaller $\sigma_{\rm obs}$ compared to their immediate 
surrounding. This is probably because the surrounding gas is 
affected by the protostellar activity (e.g., molecular outflows)  
for some cores. For instance, there is a molecular outflow 
emanating from \#2 in the immediate vicinity of cores \#4 and \#24,   
thus the H$^{13}$CO$^{+}$ line widths around both \#4 and \#24 
can be broadened by this outflow activity (see Figure~\ref{fig:cont}). 
The outflow driven by core \#8 also affects the molecular gas 
around the core \#4. The details of molecular outflow analysis 
is beyond the scope of this paper but is the topic of a followup 
paper.

Furthermore, some of the cores indeed have narrow line 
widths as revealed by the NH$_{2}$D line, which is less affected 
by the molecular outflows than H$^{13}$CO$^{+}$ as mentioned 
above. In addition, there are no outflow signatures around these 
cores. For instance, the observed velocity dispersion appears to 
decrease toward the center of M1 (See Figure~\ref{fig:Rsig}). 
A trend of $\sigma_{\rm obs}$ decreasing with decreasing radial 
distance ($R_{\rm dist}$) from the center of core is found in 16 
cores (see Figure~\ref{fig:Rsig}), including 6 continuum cores 
(\#5, \#10, \#22, \#25, \#35, and \#37) and 
10 NH$_{2}$D cores (M1, M4, M6, M7, M8, M9, M13, M10, M15, 
and M17; see also Figure~3 in Paper~II for M1). 
Note that the annularly averaged $\sigma_{\rm obs}$ has relatively 
large uncertainties toward the outer edges of the cores due to the 
low S/N. The decreasing trend of $\sigma_{\rm obs}$ toward these  
core centres may indicate that turbulent dissipation from the 
filaments to the embedded objects is ongoing, enabling  
the dense precursors to collapse to form protostars. 
Alternatively, a number of of theoretical studies 
suggest that for pre-stellar cores the line width will be 
smaller in the more central regions if the infall speed 
decreases toward the center because of an outside-in collapse 
\cite[e.g.,][]{1985MNRAS.214....1W,2000ApJ...540..946L,
2021MNRAS.502.4963G}.  
In summary, some dense cores indeed have narrower 
observed velocity dispersion compared to their natal filaments. 
This may indicate that turbulent dissipation is taking place  
in these embedded cores.

\subsection{Filament Stability}
\label{sec:stab}
The comparison between the $M_{\rm line}$ and 
the corresponding critical line-mass 
$M_{\rm crit} = 2\sigma_{\rm eff}^{2}/G$ can be used to 
evaluate the stability of the filament; where $\sigma_{\rm eff}$ 
is the effective velocity dispersion and $G$ is the gravitational 
constant (see Appendix~\ref{ap:mcrit} for the estimation of critical 
line-mass.). 
Ignoring external pressure and magnetic fields, 
we computed the $M_{\rm crit}$ for thermally supported 
($\sigma_{\rm eff}=c_{\rm s}$), nonthermal motions supported 
($\sigma_{\rm eff}=\sigma_{\rm nth}$), and total motions supported  
(i.e. including both thermal and nonthermal contributions, 
$\sigma_{\rm eff}=\sqrt{c_{\rm s}^{2} + \sigma_{\rm nth}^{2}}$) 
filaments.  
As shown in Figure~\ref{fig:mass}, $M_{\rm line}$ is larger than 
the thermal critical mass ($M_{\rm crit, th}$), except for F4b, 
F6 and F13 that are smaller than the $M_{\rm crit, th}$. 
This indicates that the filaments would be gravitationally bound  
(except for F4b, F6 and F13) in the purely thermally supported case.  
$M_{\rm line}$ is about 2 times the nonthermal critical mass 
($M_{\rm crit, nth}$), which suggests that nonthermal support 
alone cannot prevent gravitational collapse.  The ratios of 
$M_{\rm crit, nth}$/$M_{\rm crit, th}$ are in the range 0.4--1.4 
with a mean value of 0.7, which suggests that  the filaments 
are mostly supported by thermal motions. 
The estimated $M_{\rm line}$ is smaller than the total critical 
($M_{\rm crit, tot}$) mass in all the filaments, except for F4.

Although most of the filaments at the current evolutionary state 
are  gravitationally unbound when considering only the balance 
between self-gravity and the thermal plus nonthermal support, 
the presence of dense cores suggests that in fact star formation 
has already started.    
Note that by neglecting external pressure, magnetic field, 
mass uncertainty, and inclination angle uncertainty  
might bring an addition error into the 
$M_{\rm line}/M_{\rm crit, tot}$. 
Being gravitationally bound is not the sole prerequisite  
for forming stars in a filament.  
The fragmentation may have occurred already very early in the 
evolution of the filaments, if these dense cores originate 
from filament fragmentation. 
In addition, the subsonic and transonic dominated filaments and 
embedded cores indicate that  there are low  
turbulence environments (Paper~II); 
this is analogous to the situation 
in  low-mass star-forming clouds 
\citep[e.g.,][]{2002ApJ...578..914H,2010ApJ...712L.116P,
2011A&A...533A..34H,2016A&A...587A..97H,2017A&A...606A.123H}.  
The similarity suggests that similar turbulent conditions 
may apply in the very early evolutionary phases of low- and 
high-mass star formation at clump scales ($\leqslant$ of a few pc) 
where turbulence inherited from larger scales (e.g., giant 
molecular clouds) has already decayed or dissipated in a short 
timeframe \citep{1999ApJ...524..169M,2004RvMP...76..125M}.

Figure~\ref{fig:mass} shows $M_{\rm line}$ as a function of 
$\sigma_{\rm nth}/c_{\rm s}$. The derived masses per unit length 
are similar to those of narrow filaments in B213-L1495 
(24$\pm$19 \Mo pc$^{-1}$), Musca (26 \Mo pc$^{-1}$), NGC\,1333 
(34$\pm$22 \Mo pc$^{-1}$), and Orion 
\citep[23$\pm$11 \Mo pc$^{-1}$ for ISF, 20$\pm$18  \Mo pc$^{-1}$ 
for OMC-1, and 26$\pm$21 \Mo pc$^{-1}$ for 
OMC-2;][]{2013A&A...554A..55H,2016A&A...587A..97H,
2017A&A...606A.123H,2018A&A...610A..77H}. 
The measured $\sigma_{\rm nth}/c_{\rm s}$ are also comparable to 
those narrow filaments in the B213-L1495, Musca, NGC\,1333, and 
Orion (OMC-1/2 and ISF; see Figure~\ref{fig:mass}).  
These results indicate that the masses per unit length and gas 
kinematics of narrow filaments in NGC\,6334S are comparable 
to those found in various other environments,  from low-mass to 
high-mass star-forming molecular clouds.

\begin{figure*}[ht!]
\epsscale{1.2}
\plotone{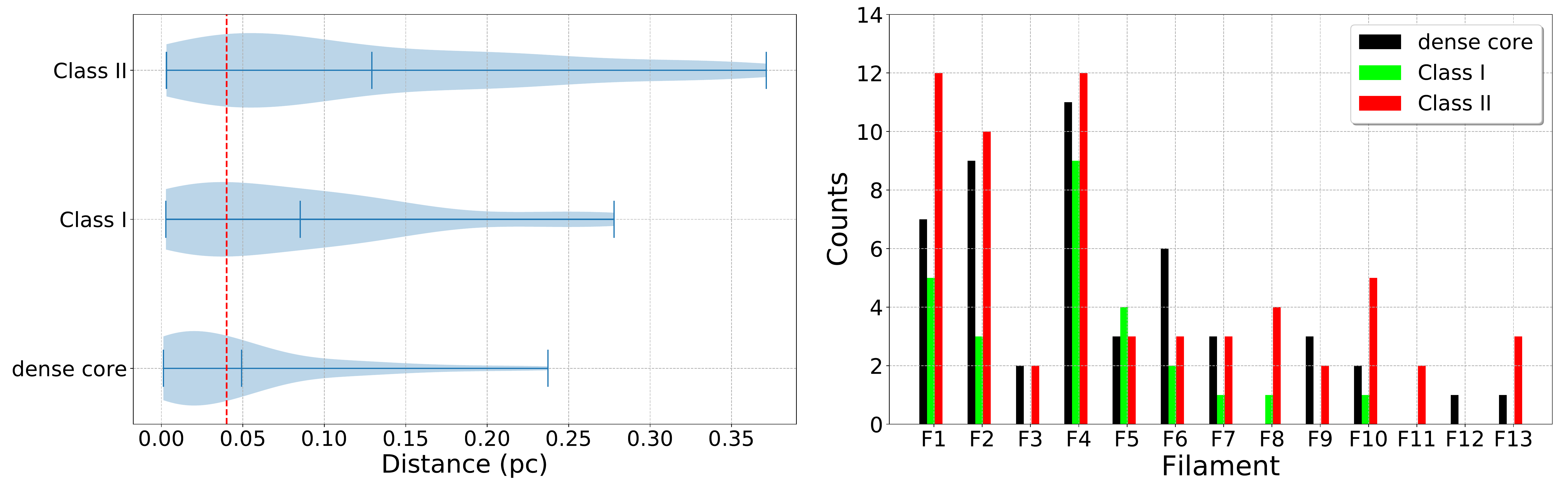}
\caption{Left: violin plot of the distance distributions for 
each type object, where the distances are the objects to the 
nearest filament spine.  The shape of each distribution shows 
the probability density of the data smoothed by a kernel 
density estimator. The blue bars from the top to bottom 
represent the maximum, mean, and minimum values, 
respectively. The vertical red dashed line is the mean 
beam-convolved filament width of $\sim$0.04~pc. 
Right:  the number of the nearest dense cores, Class~\1 
and Class~\2 objects for each filament.
\label{fig:dist}}
\end{figure*}


The total gas mass computed from H$^{13}$CO$^{+}$ is about 
395 \Mo, which is larger than the total gas mass of 160 \Mo 
estimated from continuum emission.  The H$^{13}$CO$^{+}$ 
recovers about 28\% of the accumulated mass (1389 \Mo) 
derived from H$_{2}$ column density map (derived in Paper~I; 
see Appendix A in Paper~I for detailed derivation of H$_{2}$ 
column density.)  
This indicates that the extended flux, which contains 
a significant amount of mass, is not fully recovered by the   
H$^{13}$CO$^{+}$ line toward NGC\,6334S in this ALMA 
observation.  The estimated filament masses  should 
thus be treated as lower limits because  
H$^{13}$CO$^{+}$ (1-0; $n_{\rm cr} \sim \, 10^{5}$ 
cm$^{-3}$) only probes higher density gas components; 
moreover, the data suffer from missing flux due to the lack 
of short spacing observations.

\subsection{Population of Embedded Cores and YSOs}
\label{YSOs}
The continuum cores are likely at protostellar or pre-stellar 
evolutionary phases, while the NH$_{2}$D cores seem at the 
starless/pre-stellar phases (see Paper~II). 
Figure~\ref{fig:rgb}(b) shows the distribution of identified 
dense objects and filaments toward NGC\,6334S.  The majority 
of the 45 continuum cores are associated with filaments, 
while only 4 continuum cores (\#22, \#28, \#33 and \#48) are 
not associated with any identified filament. In addition, 
15 out of 17 NH$_{2}$D cores are associated with filaments. 
These results indicate that the majority of dense cores 
are closely related with filaments in NGC\,6334S; 
as noted earlier, this situation is also found in nearby 
low-mass star-forming regions  
\citep[e.g.,][]{2010A&A...518L.102A}.

Figure~\ref{fig:rgb}(b) also shows the YSOs spatial distributions 
toward NGC\,6334S. There are 25 Class~\1 and 58 Class~\2 YSOs in 
the NGC\,6334S.  The Class~\1 and \2 YSOs are identified with the 
near-IR (NEWFIRM) and mid-IR (IRAC) data 
\citep[see][]{2013ApJ...778...96W}.  
Among the 49 identified continuum cores, 12 cores are spatially 
associated with Class~\1 objects, 5 cores are spatially associated 
Class~\2 objects, and the remaining 32 cores do not have YSOs 
counterparts. This indicates that these 32 cores could be younger 
compared to those cores associated with Class~\1 
and Class~\2 YSOs. The majority of YSOs have no continuum core 
counterparts, perhaps because their continuum emission is too 
faint  (1$\sigma\; \sim$ 0.03 mJy beam$^{-1}$, or 
$\sim$0.04 \Mo\ at a temperature of 10 K).  
Furthermore, the YSOs are not associated with  NH$_{2}$D 
cores counterparts, because the NH$_{2}$D line emission is 
in cold dense gas still in its extremely early evolutionary 
stages (e.g., starless and/or pre-stellar).

The majority of continuum cores, NH$_{2}$D cores, and Class~\1 
objects reside in or close to a filament, while the majority 
of Class~\2 objects are far away 
(see Figures~\ref{fig:rgb} and \ref{fig:cont}).  
We computed the distance of these objects to their nearest 
filament spine, in order to search for possible correlations 
between the evolutionary stages and the distance from the 
filament.  Based on the distance distribution of each type of 
object shown in Figure~\ref{fig:dist}(a), Class~\2 objects have 
larger distances  than Class~\1, while Class~\1 objects have 
larger distance than the distribution of continuum cores and 
NH$_{2}$D cores. Continuum cores and NH$_{2}$D cores are 
classified as the same type of object in this analysis because 
the majority of them are embedded in filaments and their 
evolutionary stages (pre-stellar or protostellar) are earlier 
than Class~\1/\2 (Paper~II).  
The median distances are 0.09~pc, 0.06~pc, and 0.03~pc for 
Class~\2, Class~\1, and  dense cores, respectively. Overall, 
Figure~\ref{fig:dist}(a) indicates that the more evolved objects 
are further away from the dense gas filaments in NGC\,6334S.

One possible explanation for the different distance distributions 
is that the evolved objects are moving away from their parental  
dense filament due to the kinematical motions \citep[e.g., 
slingshot mechanism and ejection;][]{2016A&A...590A...2S,
2020A&A...642A..21R}. 
Assuming the Class~\2 are moving 1 \kms relative to the filaments 
\citep[the typical moving velocity of Class~\2 in 
Orion;][]{2016A&A...590A...2S}, the estimated moving timescales 
are between 3$\times 10^{3}$ and 4$\times 10^{5}$ yr, with a 
median value of 9$\times 10^{4}$ yr. 
We would like to stress the fact that the actual moving distances 
may be much smaller than the estimated distances because the 
YSOs might not necessarily form in the centre of the filament. 
Therefore, the actual dynamical timescales could be smaller 
than the estimated values. 
Another possibility is that NGC\,6334S has experienced star 
formation before, and the parental molecular structures of 
Class~\2 have already been moved away from the YSOs 
\citep[e.g.,][]{2017MNRAS.467.1313V,2020A&A...642A..87K} or 
dispersed/destroyed by star formation feedback. 
Finally, we cannot rule out the possibility that a few 
Class~\2 objects may have originated outside of NGC\,6334S;  
especially those objects that are distant from the filaments.

The number of nearest dense cores, Class~\1, and Class~\2 for 
each filament is presented in Figure~\ref{fig:dist}(b). 
The number of dense cores and Class~\1 around F4 is much higher 
than for the rest of filaments, while the number of Class~\2 
around F1, F2, and F4 is comparable and higher than in the other 
filaments.  F4 is located at the central region where encompasses   
a significant fraction of dense gas and thus it has potential to 
form more stars as evidenced by the numerous continuum 
cores and YSOs.  F2 has the longest physical length in  
NGC\,6334S, and thus, it is expected to be associated with 
more YSOs.  As shown in Figure~\ref{fig:rgb}, a cluster of 
YSOs is forming on the western side of F1, resulting in a 
large number of nearest YSOs.  
We note that F1 is only a small part of a much larger filamentary 
structure seen in the infrared image (see Figure~\ref{fig:rgb}),  
implying that it has a large dense gas reservoir from which to 
form more stars.   

In summary, all identified filaments show a narrow width  
 and the majority of them host embedded dense core. 
These embedded dense cores are born in environments of 
low turbulence, which is similar to conditions  
found in low-mass star-forming regions.  
More evolved objects are found to be farther away from the 
filaments, suggesting YSOs or filaments have a tendency 
to move away from their natal place as they evolve.

\section{Conclusion}
\label{sec:con}
In this paper, we investigated the velocity-coherent filaments 
in the massive IRDC NGC\,6334S using ALMA observations.  
Using the H$^{13}$CO$^{+}$ (1-0) line emission, we have 
identified 13 velocity-coherent filaments. We investigated 
the physical properties of the identified filaments 
and characterized the dense objects in the NGC\,6334S. 
Our main findings are summarized as follows: 
\begin{enumerate}
  \item 
The filaments show a compact radial distribution with a 
median FWHM$_{\rm decon}$ of $\sim$0.04 pc. The derived   
filament widths are narrower than the previously proposed 
`quasi-universal' 0.1~pc filament width. 
In addition, the filament widths are roughly twice the size of 
embedded cores (radius $\sim$0.017~pc). 
The higher spatial resolution observations and higher-density gas 
tracer tend to identify even narrower and lower mass filaments.  

  \item 
The nonthermal motions are predominantly subsonic and transonic 
in all observed filaments; the single exception is F4 which 
has been significantly affected by protostellar feedback.  
The filaments are largely supported by thermal motions. 
The physical properties (mass, mass per unit length, gas kinematics, 
and width) of filaments are similar to those seen in narrow 
filaments found in various other kinds of environments such 
as low-mass, intermediate-mass, and high-mass star-forming 
regions (i.e., B213-L1495, Musca, NGC 1333, Orion, and 
G035.39-00.33).

  \item 
A fraction of the embedded objects show narrower observed 
velocity dispersions ($\sigma_{\rm obs}$) than their natal 
filaments,  which may indicate that turbulent dissipation is 
taking place in these embedded cores. 
The subsonic and transonic dominated filaments and dense 
cores indicate that in NGC\,6334S the stars are often born in 
environments of low turbulent motions. This conclusion  
hints that similar small turbulent conditions exist 
at very early evolutionary stages of low- and high-mass 
star formation at clump scales.

  \item 
The median distance to the nearest filament for dense 
cores, Class~\1, and Class~\2, is 0.03~pc, 0.06~pc, and 
0.09~pc respectively.  The increasing distances suggest 
that the more evolved objects are farther away from the 
filaments in the NGC\,6334S,  perhaps because either YSOs 
or filaments tend to move away from their natal place as 
they evolve.

\end{enumerate}

\acknowledgments
We thank the anonymous referee for the constructive report. 
D.L.~acknowledges the support from the National Natural 
Science Foundation of China grant No.~11988101.
C.W.L. is supported by the Basic Science Research Program 
through the National Research Foundation of Korea (NRF) 
funded by the Ministry of Education, Science and 
Technology (NRF-2019R1A2C1010851).
P.S. was partially supported by a Grant-in-Aid for 
Scientific Research (KAKENHI Number 18H01259) of the 
Japan Society for the Promotion of Science (JSPS). 
H.B. acknowledges support from the Deutsche 
Forschungsgemeinschaft 
(DFG, German Research Foundation) – Project-ID 138713538 – 
SFB 881 (``The Milky Way System”, subproject  B1). 
H.B. further acknowledges funding from the European Research 
Council under the Horizon 2020 Framework Program via the ERC 
Consolidator Grant CSF-648505.
J.M.G. acknowledges the support of the grants 
AYA2017-84390-C2-R and PID2020-117710GB-I00 
(AEI/FEDER, UE). 
A.P. acknowledges financial support from the UNAM-PAPIIT
IN111421 grant, the Sistema Nacional de Investigadores of
CONACyT, and from the CONACyT project number 86372 of the
`Ciencia de Frontera 2019’ program, entitled `Citlalc\'oatl: 
Amultiscale study at the new frontier of the formation and 
earlyevolution of stars and planetary systems’, M\'exico. 
I.J.-S. has received partial support from the Spanish State 
Research Agency (AEI) project number PID2019-105552RB-C41.
This paper makes use of the following ALMA data: 
ADS/JAO.ALMA\#2016.1.00951.S.
ALMA is a partnership of ESO (representing its member states), 
NSF(USA) and NINS (Japan), together with NRC (Canada)
and NSC and ASIAA (Taiwan) and KASI (Republic
of Korea), in cooperation with the Republic of Chile.
The Joint ALMA Observatory is operated by ESO,
AUI/NRAO and NAOJ.

\vspace{5mm}
\facilities{ALMA, Herschel.}

 \software{
 CASA \citep{2007ASPC..376..127M}, 
 APLpy \citep{2012ascl.soft08017R}, 
 Matplotlib \citep{4160265}, 
 Astropy \citep{2013A&A...558A..33A}, 
 PySpecKit \citep{2011ascl.soft09001G}, 
 Numpy \citep{harris2020array}.
 }


\begin{deluxetable*}{ccccccccccccccccc}
\setlength{\tabcolsep}{0.7mm}{
\tabletypesize{\scriptsize}
\rotate
\tablecolumns{12}
\tablewidth{0pc}
\tablecaption{Physical parameters of the filaments. \label{tab:fil}}
\tablehead{
&&&&&&&& \colhead{Gaussian} & &&&  &\colhead{Plummer} & 
\\ 
\cline{8-11} \cline{13-17} 
\colhead{Filament}					&\colhead{$L_{\rm fil}$}		&\colhead{$M_{\rm fil}$}		
&\colhead{$M_{\rm line}$}	
&\colhead{$M_{\rm crit,tot}$}			&\colhead{$M_{\rm crit,nth}$}	&\colhead{$M_{\rm crit,th}$}			
&\colhead{A$_{0}$}				&\colhead{stddev}			
&\colhead{FWHM} &\colhead{FWHM$_{\rm decon}$}
& &\colhead{A$_{0}$}				&\colhead{$R_{\rm flat}$}		&\colhead{$p$}
&\colhead{FWHM} &\colhead{FWHM$_{\rm decon}$}
\\ 
\colhead{}							&\colhead{(pc)}				&\colhead{($M_{\odot}$)}
&\colhead{($M_{\odot}\, \rm pc^{-1}$)}
&\colhead{($M_{\odot}\, \rm pc^{-1}$)}	&\colhead{($M_{\odot}\, \rm pc^{-1}$)}	&\colhead{($M_{\odot}\, \rm pc^{-1}$)}		
&\colhead{($10^{-2}$\,Jy/beam\,km/s)}	&\colhead{(pc)}			
&\colhead{(pc)}	 &\colhead{(pc)}	
& &\colhead{($10^{-2}$\,Jy/beam\,km/s)}	&\colhead{(pc)}				
&\colhead{}	
&\colhead{(pc)} &\colhead{(pc)}	 		
\\
\colhead{(1)}  &\colhead{(2)}     &\colhead{(3)}      &\colhead{(4)} &
\colhead{(5)}   &\colhead{(6)}    &\colhead{(7)}  
&\colhead{(8)}  &\colhead{(9)} &\colhead{(10)}  &\colhead{(11)}   
& &\colhead{(12)}  &\colhead{(13)}  &\colhead{(14)} &\colhead{(15)} &\colhead{(16)}
}
\startdata
F1 & 0.48 & 14 & 29 & 41 & 15 & 24 & 8.45$\pm$0.12 & 0.020 & 0.047 & 0.042 & &	8.53$\pm$0.12 & 0.057$\pm$0.009 & 9.78$\pm$2.37 & 0.047 & 0.042	\\  \hline
F2 & 1.37 & 52 & 38 & 47 & 20 & 24 & 5.20$\pm$0.06 & 0.026 & 0.061 & 0.057 & &	5.16$\pm$0.14 & 0.081$\pm$0.030 & 10.05$\pm$5.74 & 0.066 & 0.062	\\  \hline
F3 & 0.50 & 17 & 33 & 41 & 14 & 24 & 7.30$\pm$0.11 & 0.023 & 0.054 & 0.049 & &	7.25$\pm$0.20 & 0.051$\pm$0.012 & 6.27$\pm$1.99 & 0.056 & 0.051	\\  \hline
F4 & 1.27 & 82 & 64 & 65 & 36 & 25 & 8.39$\pm$0.23 & 0.021 & 0.049 & 0.044 & &	8.84$\pm$0.26 & 0.021$\pm$0.004 & 2.81$\pm$0.34 & 0.045 & 0.039	\\ 
F4b & 0.27 & 4 & 14 & 41 & 15 & 24 & 9.90$\pm$0.21 & 0.019 & 0.045 & 0.040 & &	10.05$\pm$0.29 & 0.014$\pm$0.003 & 1.84$\pm$0.17 & 0.057 & 0.053	\\  \hline
F5 & 0.53 & 18 & 34 & 41 & 16 & 24 & 6.37$\pm$0.16 & 0.024 & 0.056 & 0.052 & &	6.73$\pm$0.22 & 0.024$\pm$0.004 & 2.78$\pm$0.30 & 0.051 & 0.047	\\  \hline
F6 & 0.86 & 16 & 19 & 43 & 14 & 24 & 5.06$\pm$0.11 & 0.015 & 0.036 & 0.029 & &	5.16$\pm$0.15 & 0.023$\pm$0.006 & 3.94$\pm$1.08 & 0.036 & 0.029	\\  \hline
F7 & 0.62 & 15 & 24 & 40 & 15 & 24 & 3.65$\pm$0.18 & 0.016 & 0.037 & 0.030 & &	3.77$\pm$0.22 & 0.015$\pm$0.006 & 2.40$\pm$0.55 & 0.038 & 0.032	\\  \hline
F8 & 0.40 & 11 & 28 & 42 & 18 & 24 & 7.46$\pm$0.22 & 0.015 & 0.036 & 0.029 & &	7.62$\pm$0.19 & 0.034$\pm$0.007 & 7.16$\pm$2.05 & 0.034 & 0.027	\\  \hline
F9 & 0.82 & 23 & 28 & 43 & 17 & 24 & 3.92$\pm$0.13 & 0.031 & 0.074 & 0.071 & &	4.29$\pm$0.15 & 0.021$\pm$0.005 & 2.17$\pm$0.24 & 0.064 & 0.060	\\  \hline
F10 & 0.53 & 13 & 26 & 44 & 19 & 24 & 8.34$\pm$0.23 & 0.019 & 0.045 & 0.040 & &	9.09$\pm$0.25 & 0.014$\pm$0.002 & 2.29$\pm$0.19 & 0.039 & 0.033	\\ 
F10b & 0.35 & 12 & 34 & 47 & 20 & 24 & 11.09$\pm$0.23 & 0.018 & 0.042 & 0.036 & &	11.21$\pm$0.25 & 0.050$\pm$0.018 & 9.85$\pm$5.34 & 0.041 & 0.036	\\  \hline
F11 & 0.42 & 15 & 37 & 45 & 17 & 24 & 11.22$\pm$0.23 & 0.017 & 0.040 & 0.034 & &	11.35$\pm$0.25 & 0.045$\pm$0.014 & 9.11$\pm$4.18 & 0.039 & 0.033	\\  \hline
F12 & 0.43 & 18 & 42 & 47 & 18 & 24 & 7.34$\pm$0.46 & 0.019 & 0.044 & 0.039 & &	7.48$\pm$0.35 & 0.035$\pm$0.012 & 5.33$\pm$2.16 & 0.043 & 0.038	\\  \hline
F13 & 0.78 & 17 & 21 & 43 & 11 & 24 & 5.68$\pm$0.18 & 0.016 & 0.037 & 0.030 & &	5.84$\pm$0.18 & 0.030$\pm$0.007 & 5.67$\pm$1.56 & 0.035 & 0.028	\\  \hline
cont filament$^{a}$ & 0.80 &  &  &  &  &  & 0.029$\pm$0.001 & 0.016 & 0.037 & 0.032 & &	0.032$\pm$0.001 & 0.012$\pm$0.002 & 2.47$\pm$0.23 & 0.030 & 0.023	\\  \hline
mean & 0.65 & 22 & 32 & 45 & 18 & 24 & 6.84$\pm$0.18 & 0.020 & 0.046 & 0.041 & &	7.03$\pm$0.20 & 0.033$\pm$0.009 & 5.24$\pm$1.78 & 0.045 & 0.039	\\  \hline
median & 0.53 & 16 & 29 & 43 & 17 & 24 & 7.32$\pm$0.18 & 0.019 & 0.045 & 0.039 & &	7.37$\pm$0.21 & 0.027$\pm$0.006 & 4.64$\pm$1.32 & 0.042 & 0.037	\\  \hline
minimun & 0.27 & 4 & 14 & 40 & 11 & 24 & 0.03$\pm$0.001 & 0.015 & 0.036 & 0.029 & &	0.03$\pm$0.001 & 0.012$\pm$0.002 & 1.84$\pm$0.17 & 0.030 & 0.023	\\  \hline
maximum & 1.37 & 82 & 64 & 65 & 36 & 25 & 11.22$\pm$0.46 & 0.031 & 0.074 & 0.071 & &	11.35$\pm$0.35 & 0.081$\pm$0.030 & 10.05$\pm$5.74 & 0.066 & 0.062	\\  \hline
\enddata
\tablenotetext{}{Notes. 
(2) Filament length.
(3) Filament mass.
(4) Filament mass per unit length.
(5) Total critical line-mass.  
(6) Nonthermal critical line-mass.  
(7) Thermal critical line-mass.  
(8) -- (11)  The amplitude, standard deviation 
$\sigma\, = \, \rm FWHM/(2\sqrt{2ln(2)})$, and 
width, and beam-deconvolved width  derived from the 
Gaussian fitting. 
(12)--(16) The amplitude, flattening radius, density profile,  
width, and beam-deconvolved width derived from the 
Plummer fitting. 
a: the amplitude unit is Jy beam$^{-1}$. 
}
}
\end{deluxetable*}

\begin{figure*}
\center
 \includegraphics[scale=0.37,angle=90]{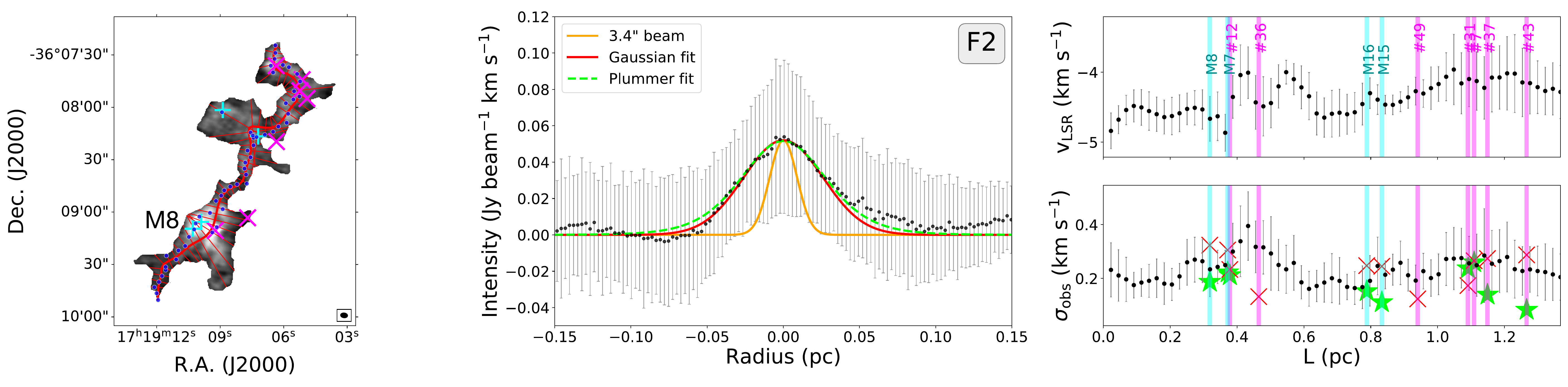}
\caption{Left column: the filament spine (red solid curve)  
overlaid on the velocity-integrated intensity image. 
Magenta cross ``x" and cyan plus ``+" symbols  are continuum cores 
and NH$_{2}$D cores, respectively.  
Middle column: mean integrated intensity profile and best-fit result  
(black dots) built by sampling radial cuts (short red solid lines) 
every 7 or 8 pixels (3\arcsec.44 corresponds to $\sim$0.019 pc at the 
source distance of 1.3 kpc) along the spine. The radial 
distance is the projected distance from the peak emission at a 
given cut (blue dots in the left column). The error bar represents 
the standard deviation of the cuts at each radial distance. 
The orange solid line shows the beam response with a FWHM of 
$\sim$3.\arcsec4 . The red solid and green dashed lines present the 
best-fit results of Gaussian and Plummer fitting, respectively.  
Right column: the mean $v_{\rm LSR}$ and mean 
$\sigma_{\rm obs}$ of H$^{13}$CO$^{+}$ line emission 
variation along the filament. The error bars show the standard 
deviation of corresponding $v_{\rm LSR}$ and $\sigma_{\rm obs}$. 
Vertical magenta and cyan lines indicate the positions of associated 
continuum cores and NH$_{2}$D cores, respectively.
The red cross ``x"  and green filled star symbols mark the core 
mean $\sigma_{\rm obs}$ derived from the H$^{13}$CO$^{+}$ and 
NH$_{2}$D lines, respectively. 
}
\label{fig:comb}
\end{figure*}

\begin{figure*}\ContinuedFloat 
\center
 \includegraphics[scale=0.42,angle=90]{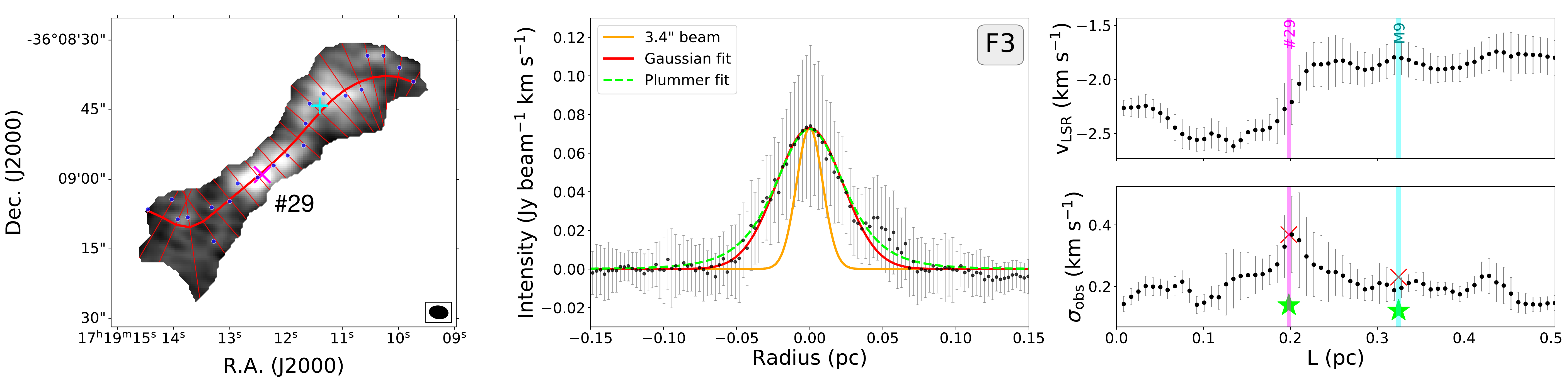}
 \includegraphics[scale=0.42,angle=90]{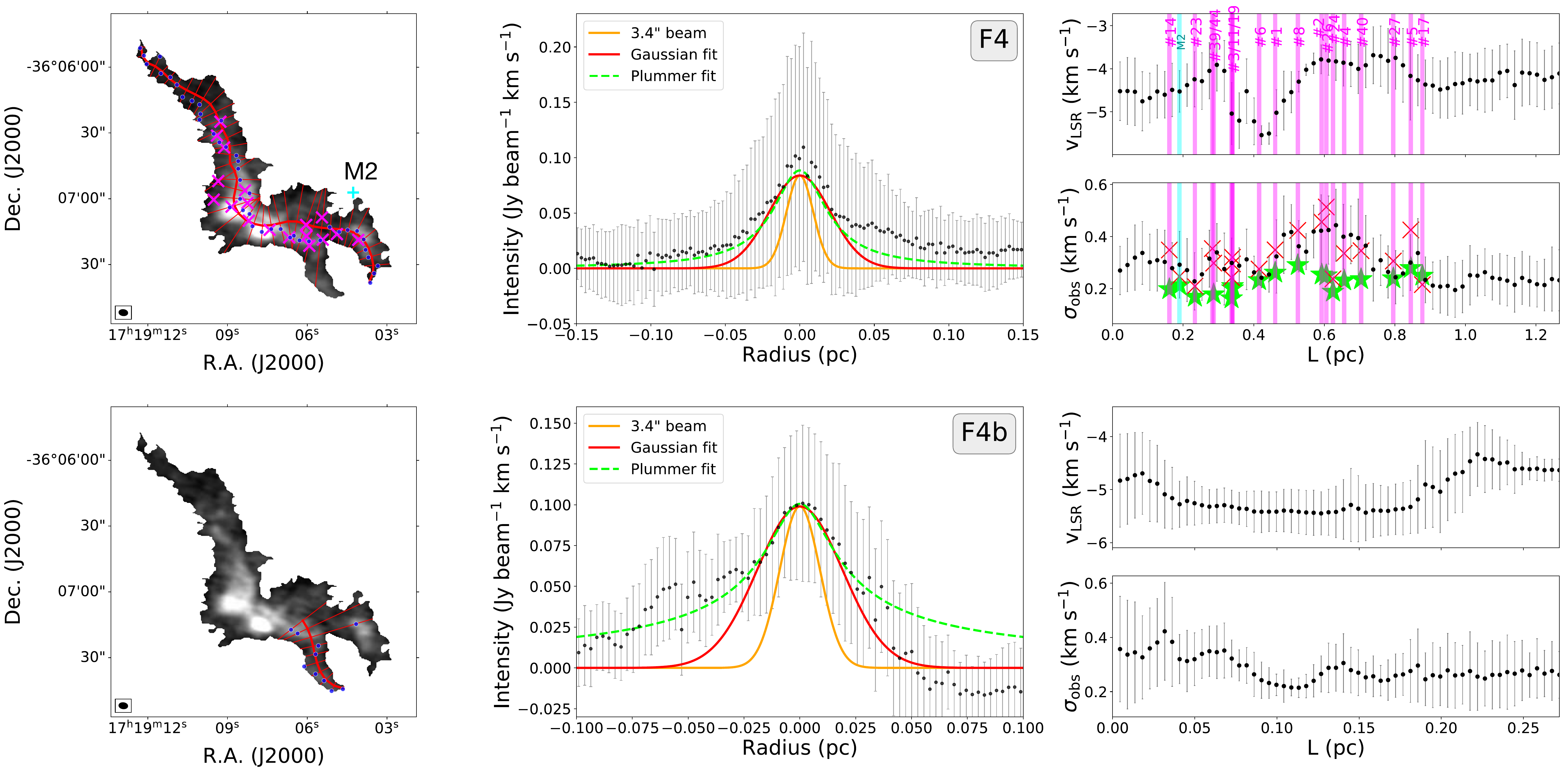}
\caption{
}
\end{figure*}

\begin{figure*}\ContinuedFloat 
\center
 \includegraphics[scale=0.42,angle=90]{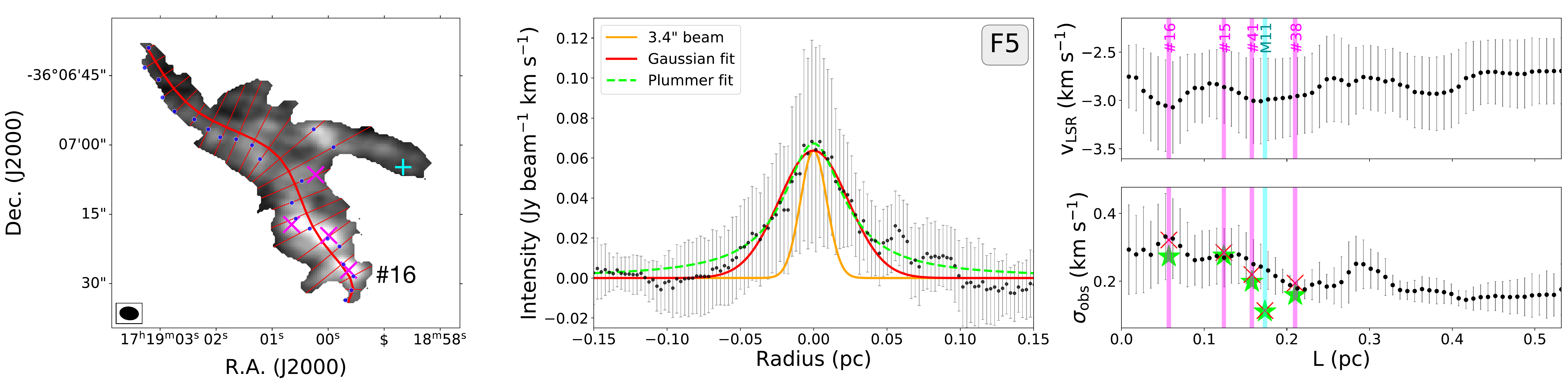}
 \includegraphics[scale=0.42,angle=90]{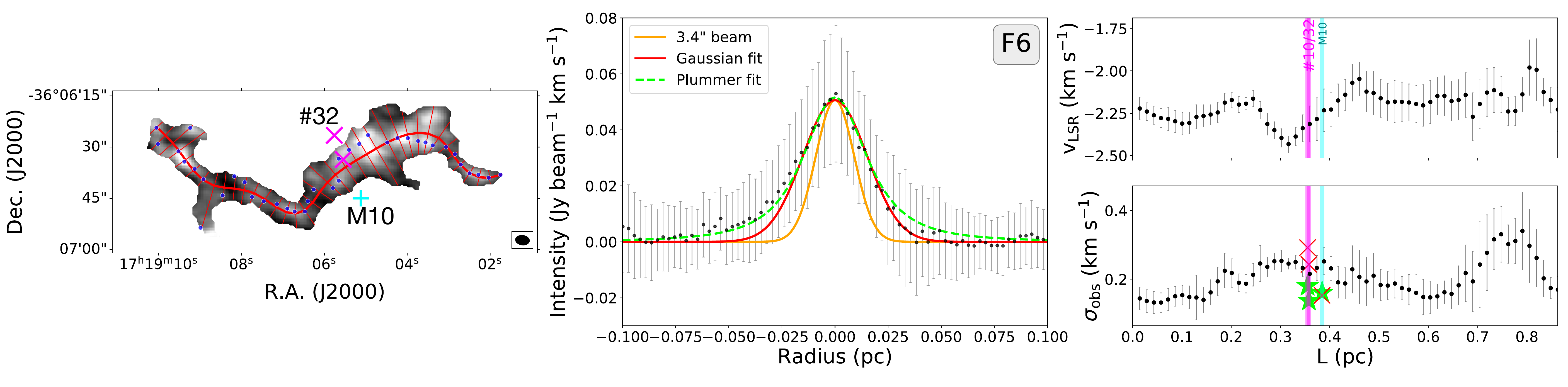}
 \includegraphics[scale=0.42,angle=90]{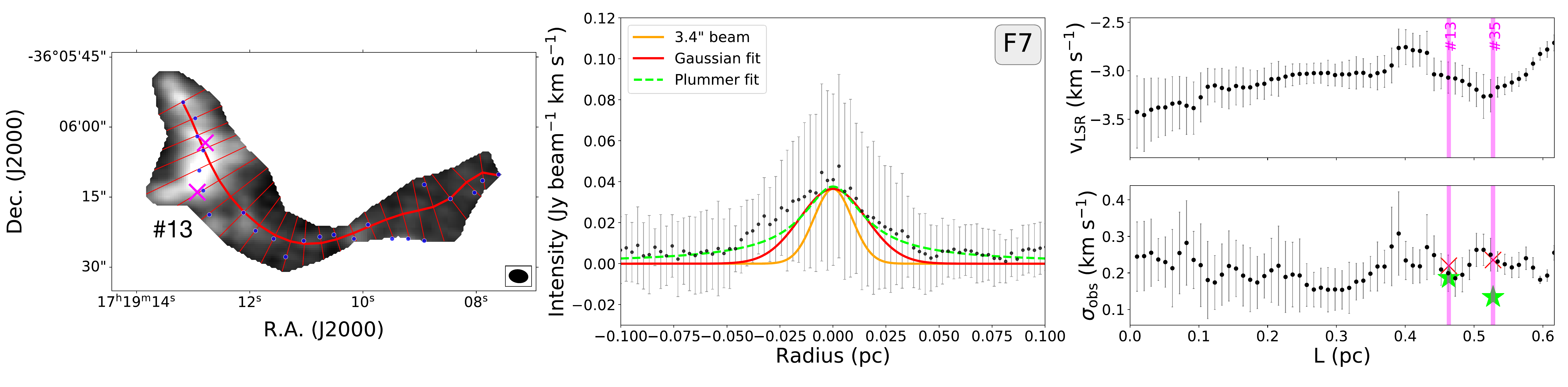}
\caption{
}
\end{figure*}

\begin{figure*}\ContinuedFloat 
\center
 \includegraphics[scale=0.42,angle=90]{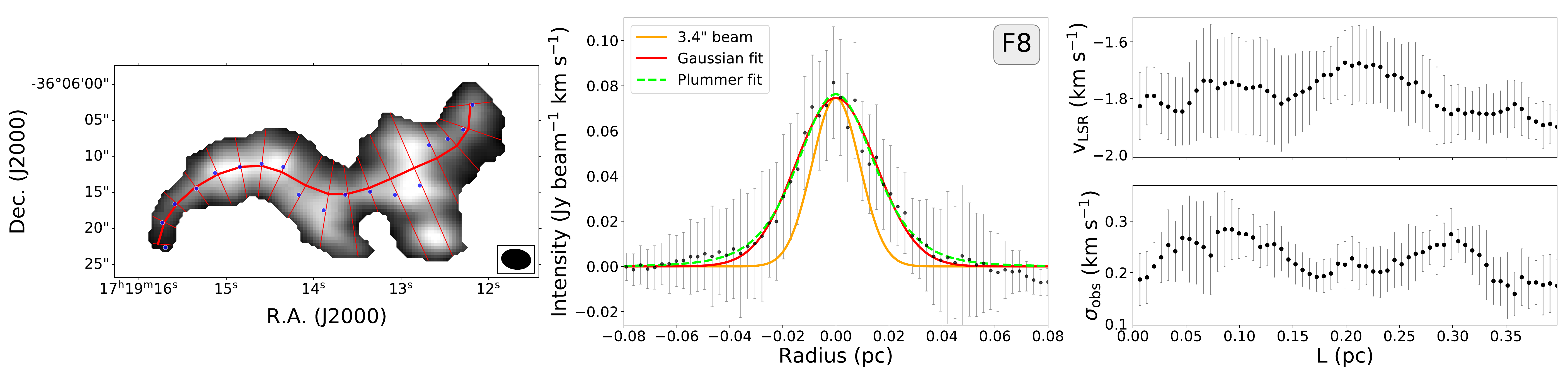}
 \includegraphics[scale=0.42,angle=90]{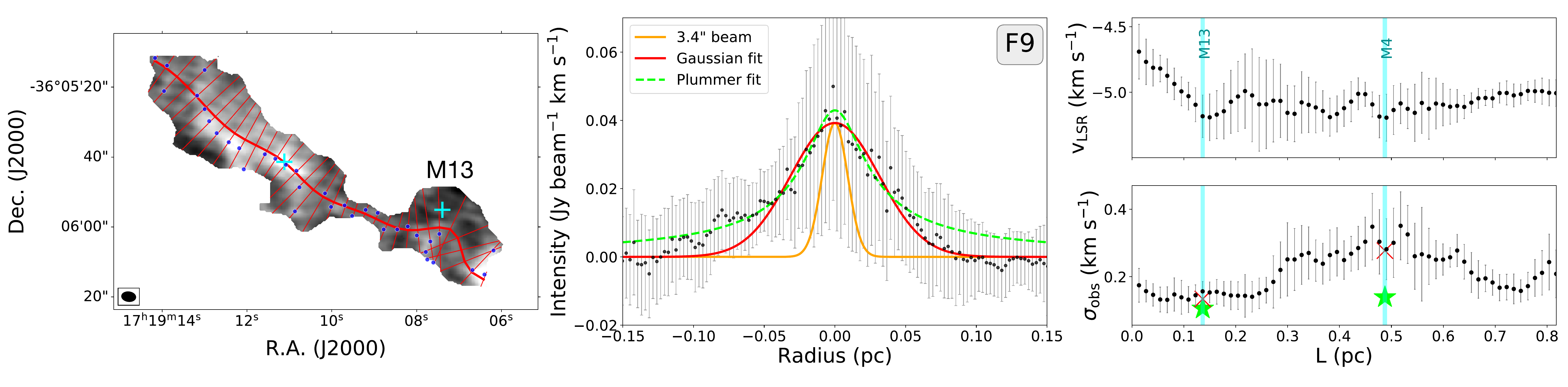}
\caption{
}
\end{figure*}

\begin{figure*}\ContinuedFloat 
\center
 \includegraphics[scale=0.42,angle=90]{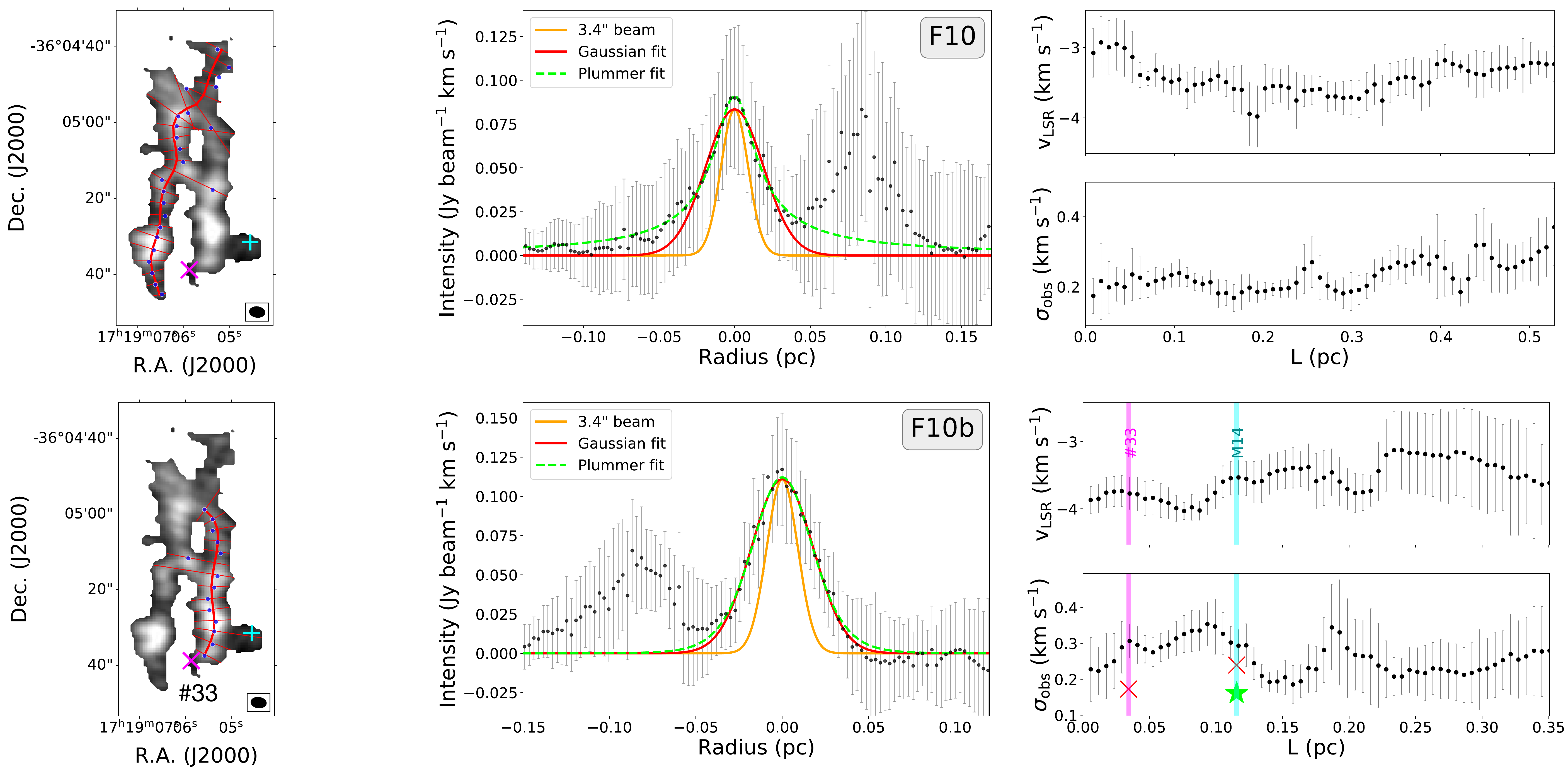}
 \includegraphics[scale=0.42,angle=90]{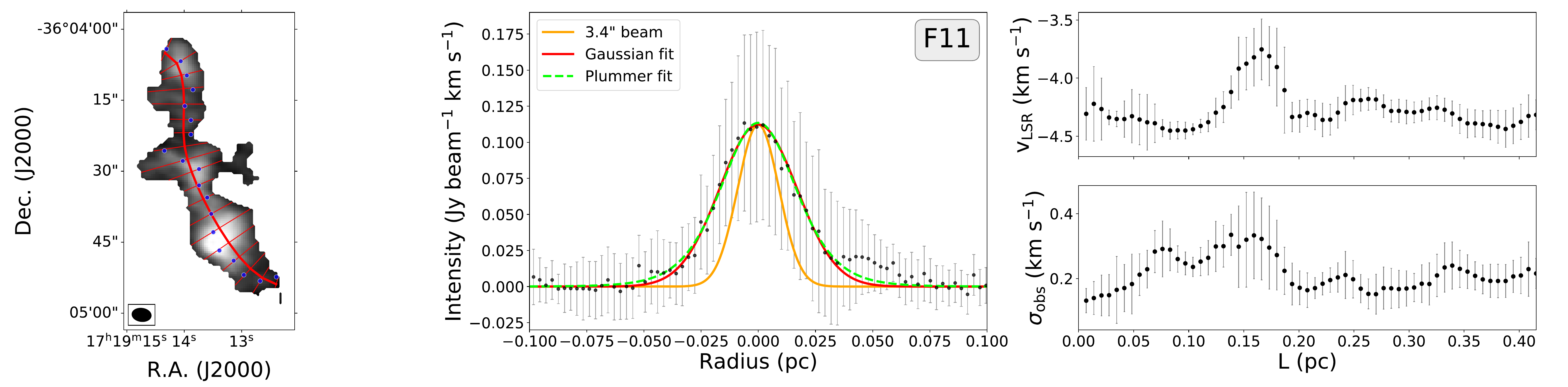}
\caption{
}
\end{figure*}

\begin{figure*}\ContinuedFloat 
\center
 \includegraphics[scale=0.42,angle=90]{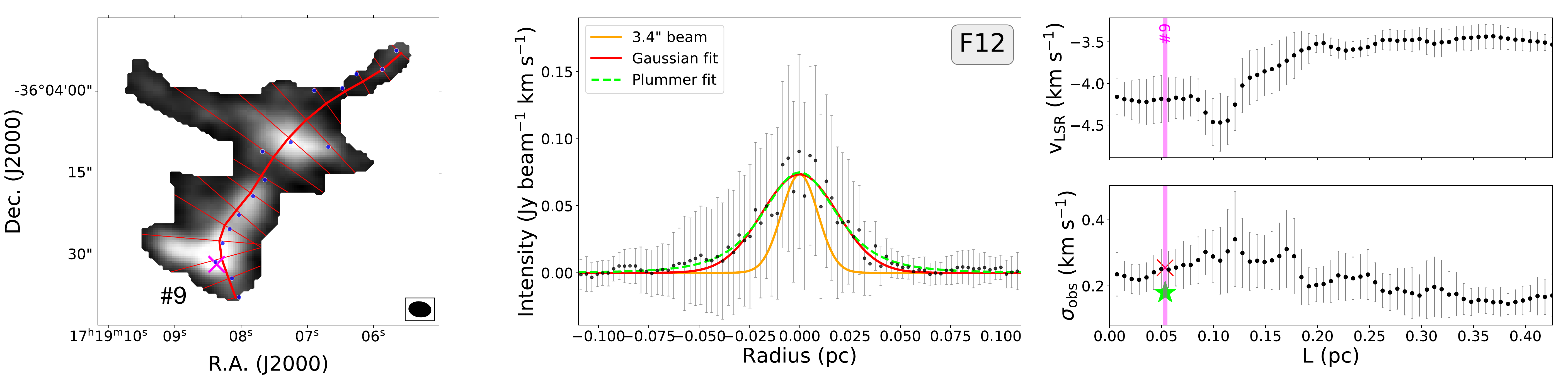}
 \includegraphics[scale=0.42,angle=90]{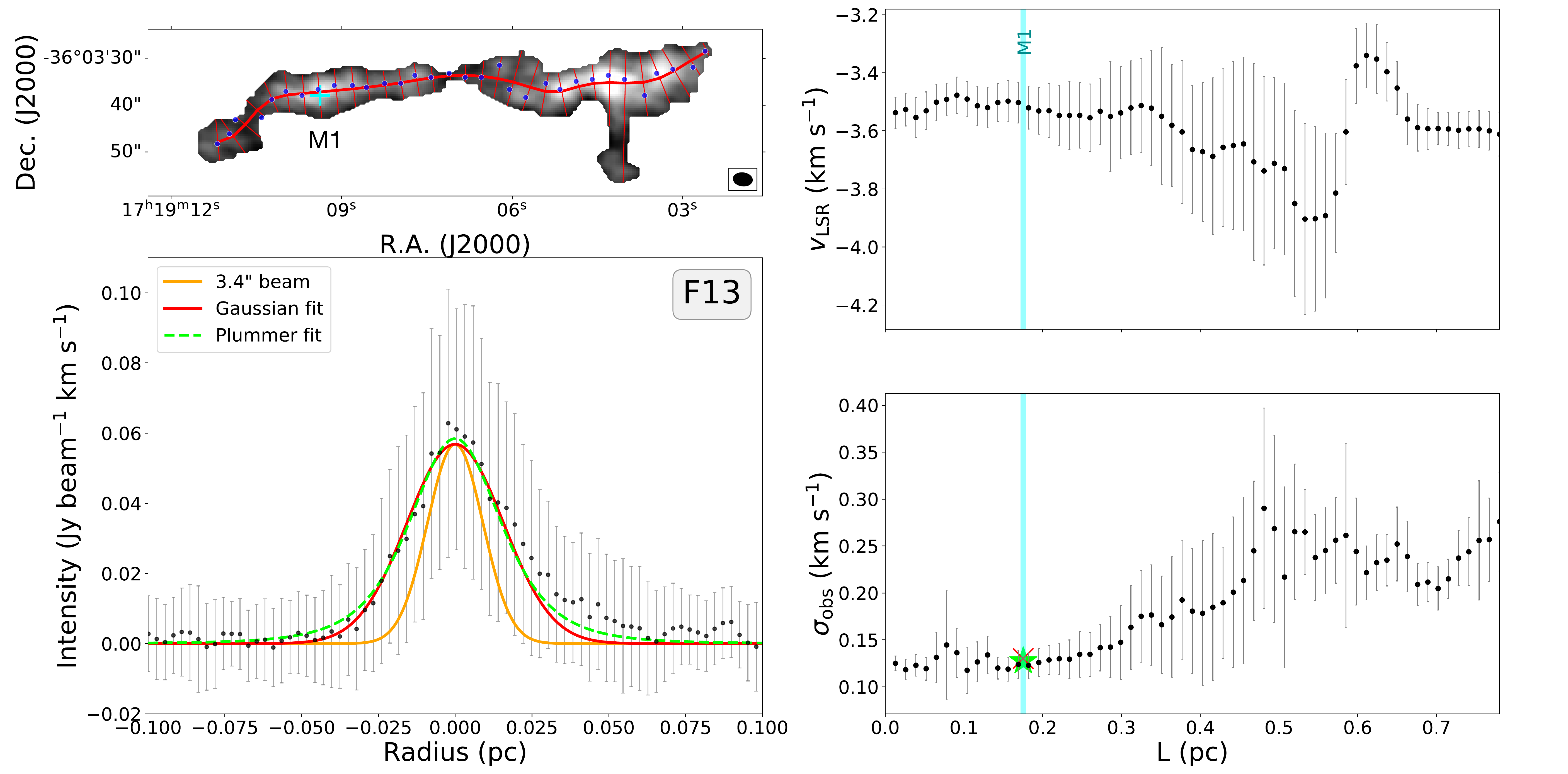}
\caption{
}
\end{figure*}

\clearpage
\bibliography{filament}{}
\bibliographystyle{aasjournal}

\appendix

\section{Column density}
\label{ap:column}
Assuming local thermodynamic equilibrium (LTE), 
the column density of molecules can be calculated  following 
\citep{2015PASP..127..266M}
\begin{equation}
\label{column}
N = C_{\tau} \frac{3h}{8 \pi^{3}  R }  
\frac{Q_{\rm rot}}{S \mu^{2} g_{\rm u}} 
\frac{{\rm exp}(\frac{ E_{\rm u} }{ k T_{\rm ex} }) }{ {\rm exp}(\frac{ h\nu }{k T_{\rm ex} }) \, - \,1}
\left(J_{\nu}(T_{\rm ex}) \, - \, J_{\nu}(T_{\rm bg}) \right) ^{-1}
\int\frac{T_{\rm R}dv}{f},
\end{equation}
where $C_{\tau}\, = \, \tau/(1 - {\rm exp}(-\tau)$  is the optical depth 
correction factor, $h$ is the Planck constant, 
$S \mu^{2}$ is the line strength multiplied by the square of dipole moment, 
$R$  is the line intensity, 
$g_{\rm u}$  is the statistical weight of the upper level, 
$T_{\rm ex}$ is the excitation temperature, 
$T_{\rm bg}$ is the back ground temperature, 
$E_{\rm u}$  is the energy of the upper state, 
$\nu$  is the transition frequency, 
$\int T_{\rm R}dv$ is the velocity-integrated intensity, 
$f$ is the filling factor, and $Q_{\rm rot}$ is the partition 
function. 
Here $f$ is assumed to be 1 and the $T_{\rm NH_3}$ approximates 
the $T_{\rm ex}$ of molecular lines (see Section~\ref{sec:oview}). 
Both H$^{13}$CO$^{+}$ and 
NH$_{2}$D emission are generally optically thin. 
The NH$_{2}$D partition function is 
$Q_{\rm rot}$ = 0.73$T_{\rm ex}^{3/2}$ + 6.56  
that is the best-fit result from a fit to the partition function obtained 
from CDMS catalogues at the different excitation temperatures of 
10--300 K \citep{2005JMoSt.742..215M}, while the H$^{13}$CO$^{+}$ 
partition function can be estimated from 
$Q_{\rm rot} \approx\, k T_{\rm ex}/h B$ +1/3 that is a approximation 
for diatomic linear molecules \citep{2015PASP..127..266M}. 
For NH$_{2}$D, the molecular parameters are 15 for $g_{\rm u}$; 
11.91 D for $S \mu^{2}$; 20.68 K for $E_{\rm u}$;  85.926 GHz for $\nu$; 
1/2 for $R$ that is the relative intensity of the main hyperfine transition 
with respect to the other hyperfine transitions. 
For H$^{13}$CO$^{+}$, the molecular parameters are 3 for $g_{\rm u}$; 
15.21~D$^{2}$ for $S \mu^{2}$; 15.21 K for $E_{\rm u}/k$;    86.754288 GHz for 
$\nu$; 1 for $R$.

The $N_{\rm H_{2}}$ is derived from the continuum emission with 
\begin{equation}
\label{N_H2}
N_{\rm H_{2}} = \eta \frac{S_{\nu}}{\Omega \, B_{\nu}(T_{\rm dust}) \, \kappa_{\nu} \,  \mu \, m_{\rm H}},
\end{equation}
where $\eta$=100 is the gas-to-dust ratio, $S_{\nu}$ is the peak 
flux density, $\Omega$ is the beam solid angle, $m_{\rm H}$ is 
the proton mass, $\mu$=2.8 is the mean molecular 
weight of the interstellar medium \citep{2008A&A...487..993K}, 
and $\kappa_{\nu}$ is the dust opacity at a frequency of $\nu$. 
We used $\kappa_{\nu}$ = 0.235 cm$^{-2}$ g$^{-1}$ by assuming 
$\kappa_{\nu}\, = \, 10(\nu/1.2 \rm THz)^{\beta}$ cm$^{-2}$ g$^{-1}$
and $\beta$ = 1.5 \citep{1983QJRAS..24..267H}.

\section{Filament Critical line-mass}
\label{ap:mcrit}
Assuming the filament is an infinite self-gravitating 
isothermal cylinder in hydrostatic equilibrium,  
the critical line-mass of filament can be estimated by \citep{1964ApJ...140.1056O}
\begin{equation}
\label{mcrit}
M_{\rm crit} = \frac{2\sigma_{\rm eff}^{2}}{G},
\end{equation}
where $\sigma_{\rm eff}$ is the effective velocity dispersion and 
$G$ is the gravitational constant. 
If the thermal gas pressure is the only force opposing gravity, the 
$\sigma_{\rm eff} = c_{\rm s}$.
If the turbulence is the only force against gravity,  
$\sigma_{\rm eff} = \sigma_{\rm nth}$.
If both thermal and turbulence supports are considered,  
 $\sigma_{\rm eff} = \sqrt{\sigma_{\rm nth}^2 + c_{\rm s}^2}$. 
In the last case, the Equation~\ref{mcrit} can be written as 
\citep[see also][]{2018A&A...610A..77H}: 
\begin{equation}
\label{mcrit1}
M_{\rm crit}(T, \sigma_{\rm nth}) = \frac{2\,c_{\rm s}^{2}}{G} \left( 1 + \left(\frac{\sigma_{\rm nth}}{c_{\rm s}}\right)^{2} \right).
\end{equation}


\end{document}